\documentclass[11pt]{article}

\usepackage[margin=1.0in]{geometry}
\usepackage{amsmath,amsthm,amssymb}
\usepackage{mathtools}
\usepackage{graphicx}
\usepackage{xcolor}                       
\usepackage{setspace}
\usepackage{multirow}
\usepackage{booktabs}
\usepackage{comment}
\usepackage{dsfont}
\usepackage{authblk}
\usepackage{lineno}
\usepackage{url}
\usepackage{natbib}
\usepackage{tabularx}
\usepackage{cancel}
\usepackage{supertabular}
\usepackage[flushleft]{threeparttable}   
\usepackage{caption}
\def\T{{\mathrm{\scriptscriptstyle T}}}
\def\spacingset#1{\renewcommand{\baselinestretch}{#1}\small\normalsize}
\defcitealias{NCEI:2023}{NCEI}
\defcitealias{IPCC:2023}{IPCC}

\bibpunct{(}{)}{;}{a}{}{,}

\spacingset{1.24}                     

\title{\textbf{Seasonal trend assessment of US extreme precipitation via changepoint segmentation}\vspace{6pt}}

\author[1]{Jaechoul Lee\thanks{Correspondence to: Jaechoul Lee. E-mail: jaechoul.lee@nau.edu}}
\author[2]{Mintaek Lee}
\author[3]{Thea Sukianto}

\affil[1]{Department of Mathematics and Statistics, Northern Arizona University, Flagstaff, AZ 86011, USA}
\affil[2]{Quora, Inc., Mountain View, CA 94041, USA}
\affil[3]{Department of Statistics and Data Science, Carnegie Mellon University, Pittsburgh, PA 15213, USA}

\date{December 3, 2025}                

\begin{document}
\maketitle

\begin{abstract}
Most climate trend studies analyze long-term trends as a proxy for climate dynamics. However, when examining seasonal data, it is unrealistic to assume that long-term trends remain consistent across all seasons. Instead, each season likely experiences distinct trends. Additionally, seasonal climate time series, such as seasonal maximum precipitation, often exhibit nonstationarities, including periodicities and location shifts. Failure to rigorously account for these features in modeling may lead to inaccurate trend estimates. This study quantifies seasonal trends in the contiguous United States’ seasonal maximum precipitation series while addressing these nonstationarities. To ensure accurate trend estimation, we identify changepoints where the seasonal maximum precipitation shifts due to factors like measurement device changes, observer differences, or location moves. We employ a penalized likelihood method to estimate multiple changepoints, incorporating a generalized extreme value distribution with periodic features. A genetic algorithm based search algorithm efficiently explores the vast space of potential changepoints in both number and timing. Additionally, we compute seasonal return levels for extreme precipitation. Our methods are illustrated using two selected stations, and the results for the US are summarized through maps. We find that seasonal trends vary more when changepoints are considered than in studies that ignore them. Our findings also reveal distinct regional and seasonal patterns, with increasing trends more prevalent during fall in the South and along the East Coast when changepoints are accounted for.
\end{abstract}

\vspace{11pt}

\newpage
\spacingset{1.29}

\section{Introduction}\label{s:intro}

Extreme weather and climate events have become more frequent and severe over time, ranging from record-breaking temperatures to flooding and landslides, both often triggered by extreme precipitation. These precipitation events are particularly critical due to their profound physical and financial impacts on human communities. According to the National Oceanic and Atmospheric Administration (NOAA)’s National Centers for Environmental Information (\citetalias{NCEI:2023}, \citeyear{NCEI:2023}), over the 30-year period of 1980--2009, floods, severe storms, and tropical cyclones caused an average of 15, 31, and 98 deaths per year in the United States (US) and its territories, with annual financial damages averaging \$3.7 billion, \$4.0 billion, and \$19.5 billion per year, respectively. In contrast, in the recent period of 2010--2022, the annual averages rose sharply to 20, 82, and 303 deaths, while financial damages increased to \$5.8 billion, \$21.3 billion, and \$59.4 billion per year on average. These figures highlight a substantial rise in both fatalities and economic losses from extreme precipitation events in recent years.

While extreme precipitation trends were traditionally studied less than annual average or total precipitation trends, recent research has increasingly examined these patterns across all regions, with particularly active efforts in the US and North America. On a regional scale, \citet{Powell:Keim:2015} found overall increases in both the intensity and magnitude of extreme precipitation in the Southeastern US. \citet{Risser:Wehner:2017} attributed the magnitude of Hurricane Harvey’s precipitation in the Houston, Texas area to human-induced climate change, linking greater variability in extreme precipitation to rising global carbon dioxide levels. In the Northeastern US, \citet{Howarth:Thorncroft:Bosart:2019} observed a significant increase of 0.3 mm year$^{-1}$ in the top 1\% of precipitation events, driven by increases in both frequency and magnitude.

On a continental scale, \citet{Kunkel:others:2013} identified overall upward trends in the frequency and intensity of extreme precipitation events in the Eastern US, though trends in the Western US were largely insignificant. Projecting future changes, \citet{Prein:others:2017} indicated  via climate simulations that hourly extreme precipitation will significantly increase across most of North America, assuming current weather patterns persist. Similarly, \citet{Kirchmeier-Young:Zhang:2020}’s analysis of annual maximum daily rainfall and 5-day rainfall totals in Southern Canada and the contiguous US projected more intense and frequent extreme precipitation events, with human-induced climate change playing a contributing role. Using large model ensembles based on historical heavy rainfall data and simulations, \citet{Thackeray:others:2022} also projected an increased frequency of precipitation extremes across North America.

Extreme precipitation and its impacts strongly influence various aspects of life, including human well-being, agriculture, the environment, and ecosystems. Rigorous study of extreme precipitation trends is therefore essential to understanding how these extremes have changed and how they may evolve in the future. Accurate estimation of such trends in climatological extreme time series requires consideration of three key features: changepoints, seasonality, and extreme value methods. \citep[cf.][]{Lee:Li:Lund:2014, Zhao:others:2019, Lee:Lee:2021}.

First, changepoints in precipitation observations, caused by factors such as station relocations, changes in measurement devices, observation times, or observers \citep[cf.][]{Groisman:Legates:1994, Daly:others:2007}, must be accounted for. Trend assessment methods assuming homogeneity in the mean and variance can yield severely biased trend estimates, particularly if changepoints are ignored or incorrectly identified \citep[cf.][]{Reeves:others:2007}. As further demonstrated in \citet{Lund:Reeves:2002} and \citet{Lund:others:2023}, neglecting such changes in trend assessment can produce erroneous conclusions, such as indicating a decreasing trend when extreme precipitation has actually increased, or vice versa. Moreover, because station metadata are often incomplete and not all changes affect trends, changepoints must be inferred directly from the precipitation data.

Second, while many studies focus on long-term trends across all seasons, seasonal trends provide valuable insights for accurate trend estimation, particularly for precipitation extremes. Seasonal variation in extreme precipitation is influenced by changes in weather patterns, atmospheric moisture, and local climate conditions. Although these patterns vary regionally, they typically follow broader seasonal climate dynamics. Therefore, trends can differ by season, and seasonal trends may not align with overall long-term trends \citep[cf.][]{Mallakpour:Villarini:2017, Li:others:2022}, a point that will be further demonstrated in our analysis with changepoints considered. Incorporating seasonal trends into the modeling framework allows for a more precise assessment of extreme precipitation behavior, with overall long-term trends derived from the aggregation of seasonal trend estimates.

Third, trend analysis using extreme value methods is more appropriate for studying extreme values. Specifically, the seasonal maximum series of daily precipitation follows an extreme value distribution, as the block size (the number of days in a season) is sufficiently large, aligning with the extreme value theorem, a counterpart to the Central Limit Theorem for seasonal mean series of daily precipitation records. Theoretically, averages and extremes are asymptotically independent \citep{McCormick:Qi:2000}, meaning trends in average precipitation are not necessarily associated with trends in extremes. In fact, precipitation changes can differ across different aspects, such as extremes and totals, as discussed in \citet{Donat:others:2016}. Therefore, extreme precipitation should be analyzed independently based on extreme value methods, in addition to averages or totals, to gain a comprehensive understanding of precipitation trends.

We aim to accurately quantify seasonal trends in the seasonal maxima of daily precipitation observations recorded in the contiguous US. To achieve this, we detect changepoints where the mean of extreme precipitation shifts, using an effective changepoint estimation method. A penalized likelihood approach is employed to find the optimal changepoint configuration based on the minimum description length criterion. Additionally, a genetic algorithm (GA) is implemented to efficiently search a large parameter space for all possible changepoint configurations \citep[cf.][]{Davis:Lee:Rodriguez-Yam:2006, Lu:Lund:Lee:2010, Li:Lund:2012, Lee:Li:Lund:2014, Lee:Lee:2021}. By incorporating the estimated changepoint information into our nonstationary extreme value model with seasonal features, we estimate seasonal trends in the US seasonal maximum precipitation series and calculate their long-term trends. Furthermore, return levels of seasonal extreme precipitation are calculated while considering changepoints, seasonalities, trends, and extreme value distributions.

The remainder of this paper is organized as follows. Section~\ref{s:data} describes the precipitation dataset used in our analysis and explains the data preprocessing procedures to extract our seasonal maximum precipitation data. Section~\ref{s:methods} explains our methods for estimating varying seasonal trends in seasonal maximum precipitation, including an extreme value model with varying parameters for different seasons and changepoints, the changepoint detection method, and the long-term trend and geostatistical smoothing approaches, selected for our analysis. Section~\ref{s:cases} illustrates our analysis methods for two selected US stations and carries out qualitative validation with station metadata. Section~\ref{s:USresults} summarizes our trend and return level estimation results for the contiguous US. In Section~\ref{s:conclusion}, we conclude with remarks and discussion.

\section{The US seasonal extreme precipitation data}\label{s:data}

The data for this study were sourced from the Global Historical Climatology Network Daily (GHCN-Daily) database, a comprehensive collection of data from over 100,000 land surface stations across 180 countries and territories \citep{Menne:others:2012,Menne:others:2018}. The GHCN-Daily dataset is part of a broader initiative to provide a modern and accurate climate record and includes daily observations of maximum and minimum temperature, precipitation, snowfall, and snow depth. For US data, the United States Historical Climatology Network (USHCN) data were utilized, which are gathered through stations within the US Cooperative Observing Network and the National Weather Service. These records are compiled by the National Centers for Environmental Information (NCEI) and the Carbon Dioxide Information Analysis Center (CDIAC). The resulting 1,218 stations in the USHCN dataset were selected based on criteria such as record longevity, coverage of the contiguous US, minimal missing records, and few significant station changes (e.g., new equipment or station relocation). The data underwent rigorous quality control procedures to ensure internal consistency, identify outliers, assess frequent values, and confirm spatial consistency, all conducted by the NCEI. The daily precipitation data files are available on the NCEI website at \url{https://www.ncei.noaa.gov/pub/data/ghcn/daily/hcn/}.

For our seasonal extreme precipitation study, we downloaded daily precipitation data and removed any obvious erroneous entries by ensuring that no values exceeded the record high for daily precipitation and that all values were nonnegative. The data were then aggregated and divided into four seasons: spring (March-April-May), summer (June-July-August), fall (September-October-November), and winter (December-January-February) to extract seasonal precipitation maxima. A seasonal maximum precipitation was determined only if no more than five days in that season were missing and at least 84 days had non-missing records. Otherwise, the season was treated as missing. For most stations, the processed seasonal maximum precipitation data extend through February 2024 (i.e., Winter 2023).

\begin{figure}[!ht]
\vspace{-0mm}
\centerline{\includegraphics[width=0.70\linewidth]{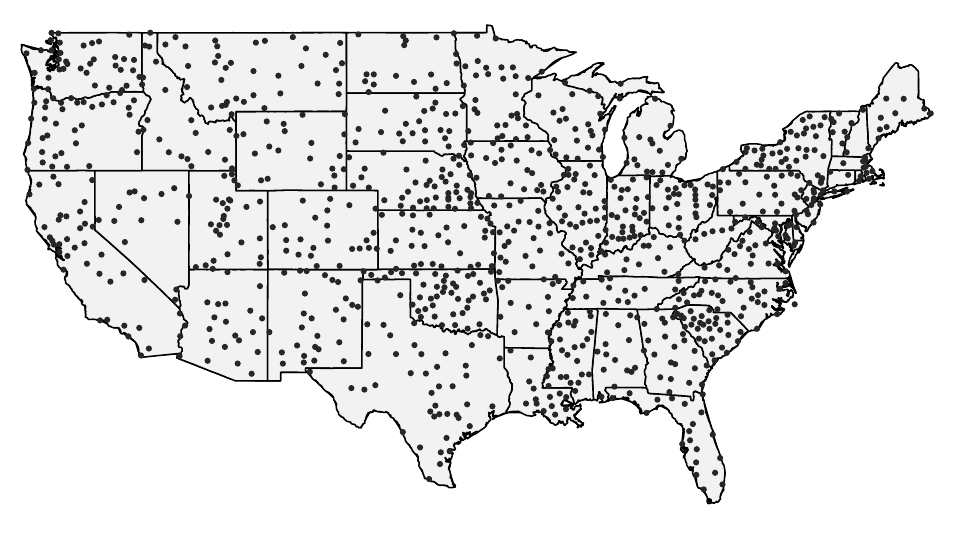}}
\vspace{-2mm}
\caption{Spatial location of the selected 1,057 USHCN stations}
\label{f:Map_station}
\end{figure}

Our study includes those stations with records spanning at least 75 years with less than 30\% missing data or 45--74 years with less than 7.5\% missing data, as similarly selected in \citet{Lee:Li:Lund:2014}. Additionally, 12 stations were excluded due to numerical issues occurred in model fitting during our preliminary data analysis steps. These data processing criteria resulted in 1,057 stations in the contiguous US. Figure~\ref{f:Map_station} displays the spatial locations of the final 1,057 USHCN stations used in our seasonal extreme precipitation analysis. Overall, the selected stations cover most of the contiguous US, with denser coverage in the Central and Eastern regions. However, relatively sparse coverage is observed in the southern California, southern Nevada, and southeastern Oregon regions.

Additionally, we used station metadata on the NCEI website at \url{https://www.ncdc.noaa.gov/cdo-web/datasets/} to evaluate whether the changepoints detected in our analysis correspond to actual physical changes in station characteristics. Although incomplete, the metadata document notable events and historical changes, including station relocations, instrumentation updates, observation time shifts, and observer changes. These changes are organized into sub-periods, rather than documenting exact change dates, indicating approximate time windows when such changes occurred. Alterations in location, elevation, measurement devices, observers, or observation schedules can affect the consistency of precipitation measurements. However, as noted in the trend analysis of monthly extreme temperatures by \cite{Lee:Li:Lund:2014}, not all recorded events cause statistically significant shifts in precipitation observations. Despite these limitations, the metadata remain a valuable reference for validating detected changepoints.


\section{Methods}\label{s:methods}

\subsection{An extreme value model with seasonal and changepoint features}\label{s:gev}

A well-established extreme value method fits the generalized extreme value (GEV) distribution to block maximum statistics. To elaborate, suppose $P_{1}, \ldots, P_{m}$ are independent and identically distributed random variables to be observed from a population with a common cumulative distribution function $F(\cdot)$. Let $X_{(m)} = \max\{ P_{1},\ldots,P_{m} \}$ denote the maximum statistic. The Fisher-Tippett-Gnedenko theorem verifies that if there exists sequences of constants $\{ a_{m} \}$ and $\{ b_{m} \}$ with $b_{m}>0$ such that
\begin{equation*}
 P\left( \frac{X_{(m)} - a_{m}}{b_{m}} \leq z \right)
 \rightarrow F_{\text{max}}(z)
\end{equation*}
as $m \rightarrow \infty$, where $F_{\rm max}(\cdot)$ is a nondegenerate cumulative distribution function, then $F_{\rm max}(\cdot)$ belongs to one of the three types of the probability distributions: Gumbel, Fr\'{e}chet, and Weibull \citep[cf.][]{Leadbetter:Lindgren:Rootzen:1983}. The three types can be incorporated to a unified cumulative distribution function:
\begin{equation*}
 F_{\text{max}}(z)
 =\exp
  \left\{
   -\left[ 1+\xi\left( \frac{z-\mu}{\sigma} \right) \right]_{+}^{-1/\xi}
  \right\},
\end{equation*}
where $x_{+}=\max\{x,0\}$. This unified probability distribution is called the GEV distribution with location $\mu \in (-\infty,\infty)$, scale $\sigma \in (0,\infty)$, and shape $\xi \in (-\infty,\infty)$ parameters. In short, a GEV distribution can be used for the block maximum statistic $X_{(m)}$ with an adequate standardization, if block size $m$ is large enough, regardless of the population distribution $F(\cdot)$ from which $P_{1},\ldots,P_{m}$ are to be observed \citep[cf.][]{Coles:2001}.

For the extreme value analysis of seasonal maximum precipitation, let $\{ P_{1}, \ldots, P_{N} \}$ denote the daily precipitation series over $N$ days. To partition into seasonal blocks of size $m$, we express this daily series as $\{ P_{1}, \ldots, P_{m}, \ldots, P_{(n-1)m+1}, \ldots, P_{nm} \}$, where $P_{(t-1)m+j}$ represents the precipitation on the $j$-th day of the $t$-th season (or block), with $t = 1, \ldots, n$ and $N = nm$. The seasonal (block) maximum precipitation for the $t$-th season, consisting of $m$ daily observations, is then defined as $X_{t} = \max \{ P_{(t-1)m+1}, \ldots, P_{tm} \}$ for $t = 1, \ldots, n$. Because the bock size $m \approx 90$ days is sufficiently large for seasonal maxima, the GEV distribution provides an appropriate asymptotic model for the seasonal maximum precipitation series $\{ X_{1}, \ldots, X_{n} \}$.

The seasonal maximum precipitation series is not stationary over time. Different seasons are influenced by distinct climate dynamics, necessitating a model that accounts for seasonal dynamics. To incorporate these varying seasonal features, we assume that the seasonal maximum precipitation series $\{ X_{1}, \ldots, X_{n} \}$ follows a nonstationary GEV distribution, denoted as GEV($\mu_{t},\sigma_{t},\xi$), where the location parameter $\mu_t$ and scale parameter $\sigma_t$ varying over seasons. Although more flexible formulations for the shape parameter $\xi$ are possible, we assume $\xi$ to be constant, due to potential numerical instability issues during maximum likelihood estimation of the shape parameter in nonstationary GEV models \citep[cf.][]{Rust:Maraun:Osborn:2009}.

For a seasonal specification of the GEV model, we reexpress the time index $t$ using a periodic notation as $t = (k - 1)T + s$, where $k=1,\ldots,d$ indicates the calendar year and $s \in \{ 1, 2, 3, 4 \}$ denotes the season, with $s = 1$ for spring, $s = 2$ for summer, $s = 3$ for fall, and $s = 4$ for winter. Here, we adopt a natural seasonal period of $T = 4$ to reflect the four meteorological seasons, and $d=\lfloor n/T \rfloor$ represents the number of complete calendar years in the seasonal maximum precipitation series. Using this periodic notation, we model
\begin{equation}
 \mu_{t}
 = \mu_{(k-1)T+s}
 = \beta_{0,s} + \delta_{t} + \beta_{1,s}\frac{t}{100T},
\label{e:gev_mu2}
\end{equation}
where $\beta_{0,s}$ and $\beta_{1,s}$ are the baseline seasonal location parameter and the seasonal linear trend, respectively, for the season $s=1,\ldots,T$. The $\delta_{t}$ term denotes location shifts, associated with changepoint-inducing events such as changes in station location, instrumentation, observation time, and observer. Assuming that $c$ changepoints have occurred at time points $\tau_1,\ldots,\tau_c$, we model
\begin{equation}
 \delta_{t}
 = \Delta_{1}1_{[\tau_1,\tau_2)}(t) + \Delta_{2}1_{[\tau_2,\tau_3)}(t) + \cdots + \Delta_{c}1_{[\tau_c,n]}(t),
\label{e:delta}
\end{equation}
where $1_{[\tau_j, \tau_{j+1})}(t)$ denotes an indicator function returning 1 if $t$ is in time interval $[\tau_j, \tau_{j+1})$ and zero otherwise. Similarly,
\begin{equation}
 \ln \sigma_{t}
 = \ln \sigma_{(k-1)T+s}
 = \lambda_{0,s}+\lambda_{1,s}\frac{t}{100T},
\label{e:gev_sigma2}
\end{equation}
where $\lambda_{0,s}$ and $\lambda_{1,s}$ denote the seasonal baseline and linear trend of log-transformed scale parameter, respectively.

The seasonal trend parameters of the GEV model, $\beta_{1,s}$ in (\ref{e:gev_mu2}) and $\lambda_{1,s}$ in (\ref{e:gev_sigma2}), have more specific interpretations. First, $\beta_{1,s}$ represents the expected change in the mean of season $s$ maximum precipitation over a 100-year period. This follows from the identity:
\begin{align*}
 &E(X_{(k+100)T+s})-E(X_{kT+s}) \\
 &= \bigg[ \mu_{(k+100)T+s}+\frac{\sigma_{(k+100)T+s}}{\xi}\{\Gamma(1-\xi)-1\} \bigg]
  - \bigg[ \mu_{kT+s}+\frac{\sigma_{kT+s}}{\xi}\{\Gamma(1-\xi)-1\} \bigg] \\
 &= \beta_{1,s}
\end{align*}
provided $\xi < 1$, $\lambda_{1,s}=0$, and no changepoints occur during the time period. Second, when $\xi < 1/2$ and $\xi \neq 0$, $\lambda_{1,s}$ characterizes the long-term change in the standard deviation of season $s$ maximum precipitation. Specifically,
\begin{align}
 \dfrac{\sqrt{Var(X_{(k+100)T+s})}}{\sqrt{Var(X_{kT+s})}}
 &= \frac{\sigma_{(k+100)T+s}\sqrt{\{\Gamma(1-2\xi)-\Gamma^2(1-\xi)\}/\xi^2}}
         {\sigma_{kT+s}\sqrt{\{\Gamma(1-2\xi)-\Gamma^2(1-\xi)\}/\xi^2}} \nonumber \\
 &= \frac{\exp\{\lambda_{0,s}+\lambda_{1,s}((k+100)T+s)/(100T)\}}
         {\exp\{\lambda_{0,s}+\lambda_{1,s}(kT+s)/(100T)\}} \nonumber \\
 &= e^{\lambda_{1,s}},
\label{e:lambda}
\end{align}
indicating that $e^{\lambda_{1,s}}$ is the multiplicative change in the standard deviation of season $s$ maximum precipitation over a century.

While changepoint models without trend components are useful for identifying abrupt shifts in precipitation records, it is important to account for gradual and cumulative influences that potentially induce persistent trends over time. Long-term climatic changes, such as rising surface temperatures, increased atmospheric moisture capacity, changes in atmospheric circulation, and Earth's increased energy imbalance, can influence the frequency and intensity of precipitation extremes \citep[cf.][]{Trenberth:others:2003, Allan:Soden:2008, Westra:others:2013, IPCC_2021_WGI}. These factors typically evolve over decades, producing gradual impact on precipitation rather than abrupt changes. Similarly, land surface modifications such as urbanization, agricultural expansion, or deforestation can influence local convective processes and surface-atmosphere interactions, leading to diffuse and time-lagged impacts on precipitation \citep[cf.][]{Pielke:others:2007, Mahmood:others:2014}. These gradual influences cannot be completely captured by changepoint models alone and can appear as a sustained linear drift in precipitation. The GEV model parameter specification in (\ref{e:gev_mu2}) flexibly captures both abrupt shifts and gradual changes in seasonal maximum precipitation series, consistent with prior studies of climate time series with changepoint consideration \citep[cf.][]{Lund:Reeves:2002, Reeves:others:2007, Wang:others:2010, Gallagher:Lund:Robbins:2013}.

\subsection{Changepoint detection via a genetic algorithm}\label{s:GA}

The GEV model parameters in Section~\ref{s:gev} can be estimated by the maximum likelihood method. If the number of changepoints $c$ and their time locations $\tau_1, \ldots, \tau_c$ are known, the GEV model parameters to be estimated are
\begin{equation*}
\boldsymbol{\theta}
 =(\boldsymbol{\beta}^{\T},\boldsymbol{\delta}^{\T},\boldsymbol{\lambda}^{\T},\xi)^{\T},
\end{equation*}
where the GEV location, changepoint mean shifts, and scale parameters are further specified as
\begin{equation*}
\boldsymbol{\beta}
 =(\beta_{0,1},\ldots,\beta_{0,T},\beta_{1,1},\ldots,\beta_{1,T})^{\T}, \quad
\boldsymbol{\delta}
 =(\Delta_1,\ldots,\Delta_c)^{\T}, \quad
\boldsymbol{\lambda}
 =(\lambda_{0,1},\ldots,\lambda_{0,T},\lambda_{1},\ldots,\lambda_{1,T})^{\T}.
\end{equation*}
The log-likelihood function for the parameters $\boldsymbol{\theta}$ is determined using the GEV distribution function $F_{\text{max}}(\cdot)$. Specifically, if the seasonal maximum precipitation $\{ X_{1},\ldots,X_{n} \}$ are independent, the log-likelihood function for $\boldsymbol{\theta}$ can be written as:
\begin{align}
\ell(\boldsymbol{\theta})
&=-\sum_{k=1}^{d}\sum_{s=1}^{T}\ln(\sigma_{(k-1)T+s})
  -(1+1/\xi)
   \sum_{k=1}^{d}\sum_{s=1}^{T}
   \ln
   \left[
    1+\xi\left(\frac{X_{(k-1)T+s}-\mu_{(k-1)T+s}}{\sigma_{(k-1)T+s}}\right)
   \right] \nonumber \\
&\quad
  -\sum_{k=1}^{d}\sum_{s=1}^{T}
   \left[
    1+\xi\left(\frac{X_{(k-1)T+s}-\mu_{(k-1)T+s}}{\sigma_{(k-1)T+s}}\right)
   \right]^{-1/\xi},
\label{e:llikl_gev}
\end{align}
if $\xi \neq 0$ and $1+\xi(X_{(k-1)T+s}-\mu_{(k-1)T+s})/\sigma_{(k-1)T+s}>0$, with $\mu_{(k-1)T+s}$ as in (\ref{e:gev_mu2}) and $\sigma_{(k-1)T+s}$ as in (\ref{e:gev_sigma2}). Because there are no closed-form expressions for the maximum likelihood estimators for GEV parameters, a numerical optimization method is required for maximum likelihood estimation. The inverse of the Hessian matrix calculated from the optimization method can be used as an estimate for the variance-covariance matrix of the maximum likelihood estimators.

However, this standard maximum likelihood estimation procedure is often not directly applicable in practice, as complete changepoint metadata are typically unavailable, station records may provide time intervals rather than exact changepoint dates, and not every changepoint necessarily results in statistically significant changes in the precipitation record. Therefore, the changepoint parameters, the number of changepoints $c$ and their associated time locations $\tau_1,\ldots,\tau_c$, must be estimated from the data.

To estimate an optimal changepoint configuration from the data, we use a penalized likelihood method with the minimum description length (MDL) as a model penalty \citep[cf.][]{Davis:Lee:Rodriguez-Yam:2006}. This method has been very successful in finding multiple mean changes in climate time series \citep[cf.][]{Lu:Lund:Lee:2010,Li:Lund:2012,Lee:Li:Lund:2014,Lee:Lee:2021}. To elaborate, the penalized log-likelihood function is expressed as
\begin{equation}
 -2\ell_{\text{opt}}(\boldsymbol{\theta}|c;\tau_1,\ldots,\tau_c)+P(c;\tau_1,\ldots,\tau_c).
\label{e:pllikl}
\end{equation}
Here, $\ell_{\text{opt}}(\boldsymbol{\theta}|c;\tau_1,\ldots,\tau_c)$ represents the optimum value of the GEV log-likelihood in (\ref{e:llikl_gev}) at a given changepoint configuration $(c;\tau_1,\ldots,\tau_c)$, and $P(c;\tau_1,\ldots,\tau_c)$ denotes the MDL penalty for the changepoint configuration $(c;\tau_1,\ldots,\tau_c)$ as expressed as
\begin{equation}
 P(c;\tau_1,\ldots,\tau_c)
 = \ln(c+1) + \frac{1}{2}\sum_{j=2}^{c+1}\ln(\tau_{j}-\tau_{j-1}) + \sum_{j=2}^{c+1}\ln\tau_{j}
\label{e:penalty_MDL}
\end{equation}
with $\tau_{c+1}=n+1$. The first term of the MDL penalty in (\ref{e:penalty_MDL}) is the penalty associated with the number of changepoints $c$, because $c$ is an unknown integer-valued parameter to be estimated. The second term is related to the penalties from $\Delta_1,\ldots,\Delta_c$, because these location shifts are real-valued parameters to be estimated from their associated segments in the time points from $\tau_{j-1}$ to $\tau_{j}$ for $j=2,\ldots,c+1$, as explained in \citet{Davis:Lee:Rodriguez-Yam:2006} and \citet{Li:Lund:2012}. If there are any missing observations occurring during the time period from $\tau_{j-1}$ to $\tau_{j}-1$, then $\tau_{j}-\tau_{j-1}$ is replaced with the number of non-missing observations during that time period. The third term is an aggregation of the penalties corresponding to the unknown time locations $\tau_1,\ldots,\tau_c$, because these time points are integer-valued parameters having their upper bounds $\tau_2,\ldots,\tau_{c+1}$, respectively. The penalized likelihood method then finds an optimal changepoint configuration to be a changepoint configuration at which the penalized log-likelihood function in (\ref{e:pllikl}) has a local minimum.

In general, there is a critical issue on changepoints estimation. Because the changepoints number $c$ and their time locations $\tau_1,\ldots,\tau_c$ are all unknown, an exhaustive search requires $\binom{n-1}{c}$ model fits for $c=0,1,\ldots,n-1$, resulting in a total of $2^{n-1}$ model fits. Therefore, an exhaustive search method is not feasible even with a moderate sample size $n$. To resolve this computation time issue, several authors, including \citet{Davis:Lee:Rodriguez-Yam:2006}, \citet{Lu:Lund:Lee:2010}, \citet{Li:Lund:2012}, \citet{Lee:Li:Lund:2014}, and \citet{Lee:Lee:2021}, used a genetic algorithm (GA) to effectively search this large space of all possible changepoint configurations. The GAs are a stochastic search technique that finds an optimal solution for a given cost function by probabilistically mimicking the three natural evolution processes: selection, crossover, and mutation. Many GA techniques have been developed for changepoint detection problems in time series data. Therefore, we selected the GA technique developed in \citet{Davis:Lee:Rodriguez-Yam:2006} and \citet{Li:Lund:2012} to detect any significant mean shifts due to changepoints that were found in the seasonal precipitation maxima. As changepoints can significantly impact trend estimation for future time periods, detecting and analyzing the changepoints is an important key to accuracy in trend quantification.

We briefly illustrate the GA technique used in our changepoint analysis. More details are illustrated in \citet{Lu:Lund:Lee:2010}, \citet{Li:Lund:2012}, and \citet{Lee:Li:Lund:2014}. To begin, we randomly generated an initial generation of $L=200$ different changepoint configurations. Then, the GA proceeds with the following steps:
\vspace{-1.5mm}
\begin{enumerate}
\itemsep-1.0mm
\item \textit{Parent selection}: Two changepoint configurations are randomly selected as a parent with a probability proportional to their penalized log-likelihood reversed ranks. A changepoint configuration with a higher rank (i.e., a smaller value of penalized log-likelihood) is more optimal and hence is more likely to be selected.
\item \textit{Offspring production}: A `child' changepoint configuration is produced in a probabilistic manner. First, all the changepoint times of the selected parent configurations are combined, and then each time remains or drops as a changepoint time with a probability of 0.5 each. Second, each of the remaining changepoint times increases by one time unit, stays the same, or decreases by one time unit with probabilities 0.3, 0.4, and 0.3, respectively, allowing a small variation in changepoint times in a probabilistic manner.
\item \textit{Offspring mutation}: Each excluded time point in the child configuration is subject to mutation with a small mutation probability of $p_{\text{mut}} = 0.05$. This mutation step helps the GA avoid premature convergence to suboptimal changepoint configurations trapped in local minima.
\item \textit{New generation}: Steps 1--3 are repeated to generate a next generation of all $L=200$ different changepoint configurations. The smallest penalized log-likelihood valued configuration in the current generation passes over to the next generation without any alternation.
\item \textit{Convergence}: Steps 1--4 are repeated until convergence. The smallest penalized log-likelihood valued configuration in the terminating generation is selected as the optimal changepoint solution.
\end{enumerate}
After the GA determined an optimal changepoint configuration, we estimate the GEV model parameters by maximizing the log-likelihood function in (\ref{e:llikl_gev}).

\subsection{Long-term trends, return levels, and geostatistical prediction}\label{s:LT_rl_intrp}

A long-term trend in an extreme characteristic of seasonal maximum precipitation can be further assessed by aggregating the seasonal trends in our GEV model. To elaborate, we define a new trend parameter $\bar{\beta}_{1}$ as the average of the seasonal trends \citep[cf.][]{Lund:Seymour:Kafadar:2001}:
\begin{equation}
 \bar{\beta}_{1} = \frac{1}{T}\sum_{s=1}^{T}\beta_{1,s}.
\label{e:beta_LT}
\end{equation}
Because each seasonal trend $\beta_{1,s}$ represents a rate of change per century, their average $\bar{\beta}_{1}$ also reflects a century-scale trend, providing a summary measure of long-term change across all seasons. To provide a practical interpretation of $\bar{\beta}_{1}$ in our seasonal maximum precipitation model, we consider the average of all seasonal maximum precipitation values of year $k=1,\ldots,d$: $\bar{X}_{k} = \frac{1}{T}\sum_{s=1}^{T}X_{(k-1)T+s}$. The expected change in the annual seasonal maximum precipitation average series $\{ \bar{X}_{1}, \ldots, \bar{X}_{d} \}$ over a 100-year period is then expressed as
\begin{equation*}
 E(\bar{X}_{k+100})-E(\bar{X}_{k})
 = \frac{1}{T}\sum_{s=1}^{T}\big[E(X_{(k+99)T+s})-E(X_{(k-1)T+s})\big]
 = \frac{1}{T}\sum_{s=1}^{T}\beta_{1,s}
 = \bar{\beta}_{1},
\end{equation*}
provided that $\xi < 1$, $\lambda_{1,s}=0$, and no changepoints had occurred for the same period. This result indicates that the parameter $\bar{\beta}_{1}$ in (\ref{e:beta_LT}) can be interpreted as the expected change in the annual averages of seasonal maximum precipitation in a century period, when the GEV scale parameter $\sigma_{(k-1)T+s}$ is a seasonal constant without trends and there had been no changepoints occurring in the same period. For this reason, we call $\bar{\beta}_{1}$ a long-term trend of annually averaged seasonal maximum precipitation series and use the new notation $\beta_{\scriptscriptstyle\text{LTA}}$ from now on to be more consistent with this interpretation.

The long-term trend $\beta_{\scriptscriptstyle\text{LTA}}$ of the annual extreme precipitation average series can be estimated by taking the average of the GEV seasonal trend estimates $\hat{\beta}_{1,1},\ldots,\hat{\beta}_{1,T}$:
\begin{equation*}
\hat{\beta}_{\scriptscriptstyle\text{LTA}} = \frac{1}{T}\sum_{s=1}^{T}\hat{\beta}_{1,s}.
\end{equation*}
The standard error for $\hat{\beta}_{\scriptscriptstyle\text{LTA}}$ is calculated by using the estimated variances and covariances of the GEV seasonal trend estimates:
\begin{equation*}
SE(\hat{\beta}_{\scriptscriptstyle\text{LTA}})
 =\sqrt{\widehat{Var}(\hat{\beta}_{\scriptscriptstyle\text{LTA}})}
 =\frac{1}{T}[\boldsymbol{1}^{\T}\hat{\boldsymbol{\Gamma}}_{\beta_{1}}\boldsymbol{1}]^{1/2},
\end{equation*}
where $\boldsymbol{1}$ is a $T$-dimensional column vector of ones, and $\hat{\boldsymbol{\Gamma}}_{\beta_{1}}$ is the estimated $T \times T$ variance-covariance matrix of the seasonal trend estimates $\hat{\boldsymbol{\beta}}_{1}=(\hat{\beta}_{1,1},\ldots,\hat{\beta}_{1,T})^{\T}$ which can be approximated by the inverse of Hessian matrix from a selected numerical optimization method.

Another important characteristic for extreme value analysis is return levels. While the seasonal and long-term trends quantify how much precipitation extremes have changed in historic data, return levels consider prediction on future precipitation extremes. Because the seasonal precipitation maximum series is nonstationary, we use the method developed by \citet{Parey:others:2007} and \citet{Parey:others:2010}, instead of the conventional return level methods, to compute the $z$-year return level $r_{z}$ for which the expected number of precipitation exceedances in next $z$ years ($T \times z$ seasons) is equal to one. Mathematically, the $z$-year season $s$ return level $r_{z,s}$ for a nonstationary seasonal maximum precipitation series is the solution to the equation
\begin{equation}
 1 = \sum_{t=t_{I}}^{t_{I}+Tz-1}(1-F_{\text{max},t}(r_{z,s}))1_{s}(t)
\end{equation}
with $T=4$. Here, $t_{I}$ denotes a starting time point for return level computation and set to be Spring 2025 for all stations, $1_{s}(t)$ is an indicator function returning 1 if time $t$ is in season $s$ and zero otherwise, and $F_{\text{max},t}(\cdot)$ represents the nonstationary GEV cumulative distribution function with time-dependent parameters $\mu_{t}$ and $\sigma_{t}$ as in (\ref{e:gev_mu2}) and (\ref{e:gev_sigma2}), and constant shape parameter $\xi$. This nonstationary return level method produces the same result as the conventional stationary return level methods if the series is stationary \citep{Parey:others:2010}.

To uncover regional patterns of underlying spatially continuous seasonal trend and return level dynamics over the contiguous US, we use a geostatistical smoothing method. Geostatistical methods incorporate the spatial correlation structure in geospatial data and, as a result, often produce more realistic spatial prediction along with uncertainty measures compared to other smoothing methods assuming no spatial correlation \citep[cf.][]{Diggle:Tawn:Moyeed:1998}. To elaborate, our geostatistical method predicts seasonal trends at unsampled locations as follows. Suppose $\hat{\beta}_{1,s}(\mathbf{s}_{i})$, the season $s$ GEV trend estimate calculated at a location $\mathbf{s}_{i} \in D \subset \mathbb{R}^{2}$ for $i=1,\ldots,N_{\text{loc}}$ with $N_{\text{loc}} = 1,057$ stations, can be modeled as
\begin{equation}
 \hat{\beta}_{1,s}(\mathbf{s}_{i}) | S(\mathbf{s}_{i}) \sim N( \beta_{1,s} + S(\mathbf{s}_{i}), \, \omega^{2} ).
\label{e:m_geostat}
\end{equation}
Here, $\beta_{1,s}$ denotes a season $s$ constant mean trend, $\omega^{2}$ is a measurement error variance, and $S(\cdot)$ refers to a zero-mean spatial Gaussian random effect field with Mat\'ern covariance function
\begin{equation}
 Cov( S(\mathbf{s}_{i}), S(\mathbf{s}_{j}) )
 = \frac{\psi^2}{2^{\nu-1}\Gamma(\nu)}
   (\kappa \| \mathbf{s}_{i} - \mathbf{s}_{j} \| )^{\nu}K_{\nu}
   (\kappa \| \mathbf{s}_{i} - \mathbf{s}_{j} \| ),
\label{e:cov_Matern}
\end{equation}
where $\psi^2$ denotes the marginal variance of the spatial field $S(\cdot)$, and $K_{\nu}(\cdot)$ represents the modified Bessel function with order $\nu > 0$ \citep[cf.][]{Cameletti:others:2013}. Next, to predict seasonal trends at unsampled locations, we implement the stochastic partial differential equation (SPDE) approach \citep{Lindgren:others:2011} to the geostatistical model in (\ref{e:m_geostat}). In addition, we use the integrated nested Laplace approximation (INLA) approach \citep[cf.][]{Rue:others:2009, Blangiardo:Cameletti:2015, Lindgren:others:2022} to compute the posterior marginals, which can provide a computational improvement compared to classical methods such as MCMC. Overall, this model-based geostatistics approach using SPDE and INLA is a computationally efficient prediction method for geostatistical data \citep[cf.][]{Lindgren:Rue:2015} and has been applied in numerous spatial data applications \citep[cf.][]{Hagan:others:2016, Moraga:others:2021}. We also used this method to predict long-term trends, seasonal variation changes, and seasonal return levels at unsampled locations for the US.

\section{Analysis illustrations}\label{s:cases}

\subsection{Changepoint segmentation via GA}\label{s:cases_cpts}

We demonstrate our changepoint detection method using seasonal maximum precipitation series from two stations: Circleville, Ohio, and Donaldsonville, Louisiana. The Circleville station is located in Pickaway County, within the Columbus, Ohio Metropolitan Statistical Area, while the Donaldsonville station is in Ascension Parish, part of the Baton Rouge, Louisiana Metropolitan Statistical Area. These stations were selected for illustration due to the significant climatic and environmental changes affecting the region. According to the American Communities Project (ACP) in 2021, Pickaway County is designated as a `red' risk for extreme rainfall, while Ascension Parish is classified as being at a `high' risk for hurricanes and extreme rainfall. Furthermore, both Ohio and Louisiana have experienced substantial climate-related disasters, including droughts and severe storms (\citetalias{NCEI:2023}, \citeyear{NCEI:2023}).

The GA changepoint detection method in Section~\ref{s:GA} was applied to identify the optimal configuration of changepoints in the seasonal maximum precipitation series at the two stations. This was achieved by minimizing the penalized log-likelihood function in (\ref{e:pllikl}), which comprises the GEV likelihood term, evaluating the model's fitness to the data for a given changepoint configuration, and the MDL penalty term, penalizing overly complex models with more model and changepoint parameters. By jointly optimizing these terms, the GA method yields changepoint estimates that balance model fit and complexity.

After applying the GA method to the seasonal maximum precipitation series at the Circleville station, two changepoints were detected: Spring 1912 and Winter 1927. To evaluate whether these changepoints correspond to actual physical or procedural changes at the station, we examined the station metadata. Table~\ref{t:meta_Circleville} summarizes key changes in station characteristics that may have influenced precipitation measurements during the time periods surrounding the two GA-estimated changepoints. The first changepoint, Spring 1912, falls within the metadata period October 1, 1911 -- November 11, 1917, when the station elevation changed and a new observer began recording precipitation. The second changepoint, Winter 1927, coincides with the metadata period June 1, 1927 -- December 31, 1929, when the station adopted a rotating schedule (RS) for observation times (OT), with evening observations generally taken during winter months and morning observations during summer months. We note that while the metadata documents broader time periods during which such changes were present, the GA method detects specific time points at which these changes most likely produced significant shifts in the precipitation record.

\vspace{1mm}
\begin{table}[!ht]
\caption{Circleville station metadata: documented changes in station characteristics with potential impacts on precipitation observations, 1895--1917 and 1927--1929}
\label{t:meta_Circleville}
\vspace{-1mm}
{\small
\begin{threeparttable}
\begin{tabular*}{1.0\textwidth}{@{\extracolsep{\fill}}lllll@{\extracolsep{\fill}}}
\toprule
 Period (yyyy.mm.dd) & \multicolumn{1}{l}{Time} & \multicolumn{1}{l}{Change} & \multicolumn{1}{l}{Observer} & \multicolumn{1}{l}{Note} \\
\midrule
 1895.01.01 -- 1911.09.30 & RS\tnote{$\ast$} & Elevation changed, new observer & S. Courtright & OT flagged \\
 1911.10.01 -- 1917.11.11 & RS\tnote{$\ast$} & Elevation changed, new observer & H. Clarke     & OT flagged \\
 1927.01.01 -- 1927.05.31 & RS  & Precip. OT -- RS                       & H. Clarke     & OT flagged \\
 1927.06.01 -- 1929.12.31 & RS  & Precip. OT -- RS                       & H. Clarke     & OT flagged \\
\bottomrule
\end{tabular*}
\begin{tablenotes}
 \item[$\ast$] This symbol indicates that the observation time is uncertain for precipitation recordings.
\end{tablenotes}
\end{threeparttable}
}
\end{table}

Location and observer changes that likely contributed to the two changepoints at the Circleville station also appear to have induced changepoints at other stations, also occurring around similar times for some nearby sites. For example, the Kenton, Ohio station, located approximately 128 km northwest of Circleville, exhibited three GA-estimated changepoints: Summer 1902, Winter 1905, and Spring 1917, as shown in Figure~S1 (middle) of the supplementary material. According to Kenton’s metadata, these changepoints generally coincide with changes in observer, both location and observer, and again both location and observer, during the periods December 1, 1902 -- October 31, 1905; August 1, 1906 -- December 31, 1907; and January 1, 1913 -- June 10, 1924, respectively.\footnote{\scriptsize Source: \url{https://www.ncdc.noaa.gov/cdo-web/datasets/GHCND/stations/GHCND:USC00334189/detail}} Notably, Circleville’s first changepoint, Spring 1912, closely precedes Kenton’s third changepoint, Spring 1917. Similarly, the Hiram, Ohio station, located about 243 km northeast of Circleville, exhibited two GA-estimated changepoints: Spring 1930 and Spring 1952, as displayed in Figure~S1 (bottom). Hiram's metadata attributes both changepoints to observer changes, corresponding to the periods December 1, 1925 -- September 6, 1927 and September 7, 1927 -- December 31, 1956.\footnote{\scriptsize Source: \url{https://www.ncdc.noaa.gov/cdo-web/datasets/GHCND/stations/GHCND:USC00333780/detail}} Circleville's second changepoint, Winter 1927, aligns closely with Hiram's first changepoint, Spring 1930.

For the Donaldsonville seasonal maximum precipitation series, the GA method detected two changepoints: Summer 1939 and Fall 1996. Table~\ref{t:meta_Donaldsonville} shows station metadata documenting changes that potentially have impacted precipitation recordings during the periods surrounding these changepoints. Until August 31, 1918, the precipitation OT was ambiguous, coded as `9079', which indicates an unclear reporting format, typically listing `700' as the observation hour, but with uncertainty for both temperature and precipitation. From September 1, 1918 to April 30, 1940, the precipitation OT was listed as 7 a.m., while the instrument height was flagged as suspect in the metadata. Between May 1 and October 20, 1940, the OT reverted to ambiguous, again accompanied by a suspect instrument height flag. From October 21, 1940 to December 31, 1941, the OT returned to 7 a.m., and a new observer took over. This series of changes in OT and instrument setup likely contributed to Donaldsonville's first changepoint, Summer 1939. On the other hand, the second changepoint, Fall 1996, appears more directly related to the installation and use of a Fischer-Porter recording rain gauge starting from July 1, 1995.\footnote{\scriptsize Source: \url{https://www.ncdc.noaa.gov/cdo-web/datasets/GHCND/stations/GHCND:USC00162534/detail}} This instrument, developed in the mid-20th century, uses a punch-tape mechanism to automatically record precipitation and has been widely adopted by NOAA for long-term rainfall data collection.


\vspace{1mm}
\begin{table}[!ht]
\caption{Donaldsonville station metadata: documented changes in station characteristics with potential impacts on precipitation observations, 1914--1941 and 1988--2003}
\label{t:meta_Donaldsonville}
\vspace{-1mm}
{\small
\begin{threeparttable}
\begin{tabular*}{1.0\textwidth}{@{\extracolsep{\fill}}lllll@{\extracolsep{\fill}}}
\toprule
 Period (yyyy.mm.dd) & \multicolumn{1}{l}{Time} & \multicolumn{1}{l}{Change} & \multicolumn{1}{l}{Observer} & \multicolumn{1}{l}{Note} \\
\midrule
 1914.07.16 -- 1918.08.31 & 07\tnote{$\ast$} & Station moved, new observer & A. Landry  & Station 0.3 mi SE  of PO, \\
                          &     &                                      &            & instrument height flagged \\
 1918.09.01 -- 1940.04.30 & 07  & Precip. OT -- 7 a.m.                 & A. Landry  & Instrument height flagged \\
 1940.05.01 -- 1940.10.20 & 07\tnote{$\ast$} & Precip. OT -- ambiguous & A. Landry  & Instrument height flagged \\
 1940.10.21 -- 1941.12.31 & 07  & Precip. OT -- 7 a.m.,                & L. Landry  &  \\
                          &     & new observer                         &            &  \\
 1988.06.01 -- 1995.06.30 & 08  &                                      & T. Hidalgo &  \\
 1995.07.01 -- 2003.03.26 & 08  & Fischer-Porter gauge used            &            &  \\
\bottomrule
\end{tabular*}
\begin{tablenotes}
 \item[$\ast$] This symbol indicates that the observation time is uncertain for precipitation recordings.
\end{tablenotes}
\end{threeparttable}
}
\end{table}

Observation time and instrument changes contributed to changepoints at other stations as well. For example, among neighboring stations with GA-estimated changepoints identified, the Liberty, Texas station, located approximately 362 km west of Donaldsonville, exhibited two GA-estimated changepoints in Fall 1940 and Summer 1946. According to station metadata, Liberty underwent changes in location, OT, and observer during January 26, 1933 -- December 31, 1949.\footnote{\scriptsize Source: \url{https://www.ncdc.noaa.gov/cdo-web/datasets/GHCND/stations/GHCND:USC00415196/detail}} Although the exact timing of these changes is unknown, Donaldsonville’s first changepoint, Summer 1939, falls within this time period. In contrast, Donaldsonville’s second changepoint, Fall 1996, is most likely associated with an instrument transition from a standard rain gauge to a Fischer-Porter rain gauge. A comparable pattern is observed at the Cuyamaca, California station, where the GA method identified a changepoint in Fall 1995, coinciding with the installation and use of a new Fischer-Porter rain gauge during July 1, 1995 -- January 23, 1998.\footnote{\scriptsize Source: \url{https://www.ncdc.noaa.gov/cdo-web/datasets/GHCND/stations/GHCND:USC00042239/detail}} Figure~S2 in the supplementary material displays the GA-estimated changepoints at Liberty (middle) and Cuyamaca (bottom).

\subsection{Seasonal and long-term trends via changepoint segmentation}\label{s:cases_strends}

Using the GA-estimated changepoints identified in Section~\ref{s:cases_cpts}, we estimate the GEV model parameters for the seasonal maximum precipitation series by numerically optimizing the log-likelihood function in (\ref{e:llikl_gev}). Table~\ref{t:par_est_cases} summarizes the GEV model parameter estimates and their standard errors for the two stations. For the Circleville station, the seasonal trend estimates are consistently higher when the two GA-estimated changepoints, Spring 1912 and Winter 1927, are incorporated into the GEV model. Specifically, the spring and winter seasons show only minimal changes in maximum precipitation relative to their standard errors, with estimated trends of 2.338 and 3.804 mm day$^{-1}$ century$^{-1}$, respectively, when changepoints are considered, compared to --0.579 and 0.228 mm day$^{-1}$ century$^{-1}$ when changepoints are ignored. Although most trend estimates are not statistically significant, the fall trend is 8.106 mm day$^{-1}$ century$^{-1}$ and significant at the 5\% level. In contrast, when changepoints are ignored, the fall trend drops to 4.696 mm day$^{-1}$ century$^{-1}$ and is not significant. This suggests that fall maximum precipitation in Circleville, OH has been increasing more rapidly than previously estimated under the no-changepoint assumption. Moreover, fall precipitation has became significantly more variable: the standard deviation of fall maximum precipitation increases by 72.8\% (i.e., $e^{0.547} - 1 \approx 0.728$) per century when changepoints are included, compared to a 53.0\% increase (i.e., $e^{0.425} - 1 \approx 0.530$) when they are ignored. The GEV shape parameter is estimated at 0.082 when changepoints are considered and 0.061 when they are not, indicating that the distribution of extreme precipitation remains right-skewed, as expected.

\begin{table}[!ht]
\caption{GEV parameter estimates and their standard errors in parentheses for the seasonal maximum precipitation series in Circleville, OH and Donaldsonville, LA (units: mm day$^{-1}$ for $\beta_{0,s}$'s and $\Delta_{j}$'s; mm day$^{-1}$ century$^{-1}$ for $\beta_{1,s}$'s)}
\label{t:par_est_cases}
\vspace{-1mm}
{\small
\begin{tabular*}{1.0\textwidth}{@{\extracolsep{\fill}}llrrrr@{\extracolsep{\fill}}}
\toprule
                &        & \multicolumn{2}{@{}c@{}}{Circleville, OH} & \multicolumn{2}{@{}c@{}}{Donaldsonville, LA} \\
\cmidrule{3-4}\cmidrule{5-6}
 Parameter      & Season & Changepoints & No changepoints & Changepoints & No changepoints \\
\midrule
$\beta_{0,s}$   & Spring &  31.165 (2.460) &  33.436 (2.399) &  ~~55.878 (4.713) &  51.053 (4.619) \\
                & Summer &  35.615 (2.862) &  38.565 (2.826) &  ~~51.440 (4.308) &  46.745 (4.179) \\
                & Fall   &  24.250 (1.875) &  26.895 (1.834) &  ~~52.857 (5.690) &  47.838 (5.406) \\
                & Winter &  23.513 (1.871) &  26.271 (1.731) &  ~~59.383 (3.517) &  54.199 (3.618) \\
$\beta_{1,s}$   & Spring & ~~2.338 (3.169) & --0.579 (2.990) &  --11.697 (8.205) &  14.080 (6.426) \\
                & Summer & ~~5.367 (4.111) & ~~1.630 (3.866) &  --17.900 (6.982) & ~~7.424 (4.988) \\
                & Fall   & ~~8.106 (3.339) & ~~4.696 (2.872) &  --12.469 (8.874) &  13.458 (6.855) \\
                & Winter & ~~3.804 (2.745) & ~~0.228 (2.403) &  --22.173 (7.321) & ~~3.938 (5.159) \\
$\Delta_1$      &        & ~~7.574 (1.746) &                 &  ~~13.887 (3.434) &                 \\
$\Delta_2$      &        & --0.672 (1.980) &                 &  ~~28.647 (5.767) &                 \\
$\lambda_{0,s}$ & Spring & ~~2.475 (0.165) & ~~2.552 (0.164) & ~~~~3.026 (0.168) & ~~3.016 (0.166) \\
                & Summer & ~~2.505 (0.171) & ~~2.587 (0.165) & ~~~~3.080 (0.197) & ~~3.038 (0.194) \\
                & Fall   & ~~2.013 (0.146) & ~~2.134 (0.144) & ~~~~3.314 (0.167) & ~~3.306 (0.165) \\
                & Winter & ~~1.925 (0.160) & ~~2.068 (0.147) & ~~~~2.649 (0.165) & ~~2.762 (0.160) \\
$\lambda_{1,s}$ & Spring & --0.172 (0.221) & --0.238 (0.219) & ~~~~0.191 (0.224) & ~~0.266 (0.217) \\
                & Summer & ~~0.162 (0.225) & ~~0.075 (0.219) & ~~--0.353 (0.250) & --0.189 (0.239) \\
                & Fall   & ~~0.547 (0.201) & ~~0.425 (0.201) & ~~--0.066 (0.218) & --0.083 (0.216) \\
                & Winter & ~~0.291 (0.216) & ~~0.152 (0.202) & ~~~~0.391 (0.211) & ~~0.299 (0.204) \\
$\xi$           &        & ~~0.082 (0.034) & ~~0.061 (0.033) & ~~~~0.210 (0.039) & ~~0.190 (0.039) \\
\bottomrule
\end{tabular*}
}
\end{table}

The seasonal trends at the Donaldsonville station also show a strong dependence on whether changepoints are considered. Without accounting for changepoints, all estimated seasonal trends are positive, whereas including the two GA-estimated changepoints, Summer 1939 and Fall 1996, results in negative trends across all seasons. Specifically, when changepoints are ignored, the spring and fall trends are 14.080 and 13.458 mm day$^{-1}$ century$^{-1}$, respectively, both statistically significant at 5\% level. However, when the changepoints are incorporated into the GEV model, these trends become negative and nonsignificant at --11.697 and --12.469 mm day$^{-1}$ century$^{-1}$. In contrast, summer and winter show positive but nonsignificant trends, 7.424 and 3.938 mm day$^{-1}$ century$^{-1}$, when changepoints are ignored, but significantly decreasing trends of --17.900 and --22.173 mm day$^{-1}$ century$^{-1}$ when changepoints are considered. For seasonal variability, unlike Circleville, the Donaldsonville station shows no significant changes. The estimated seasonal variability parameters $\hat{\lambda}_{1,s}$ are not significantly different from zero, regardless of whether changepoints are included. Similarly, the shape parameter remains stable, with values of 0.190 without changepoints and 0.210 when changepoints are considered, indicating a consistently right-skewed distribution.

\begin{figure}[!ht]
\vspace{-4mm}
\centerline{\hspace{1mm}\includegraphics[width=0.54\linewidth]{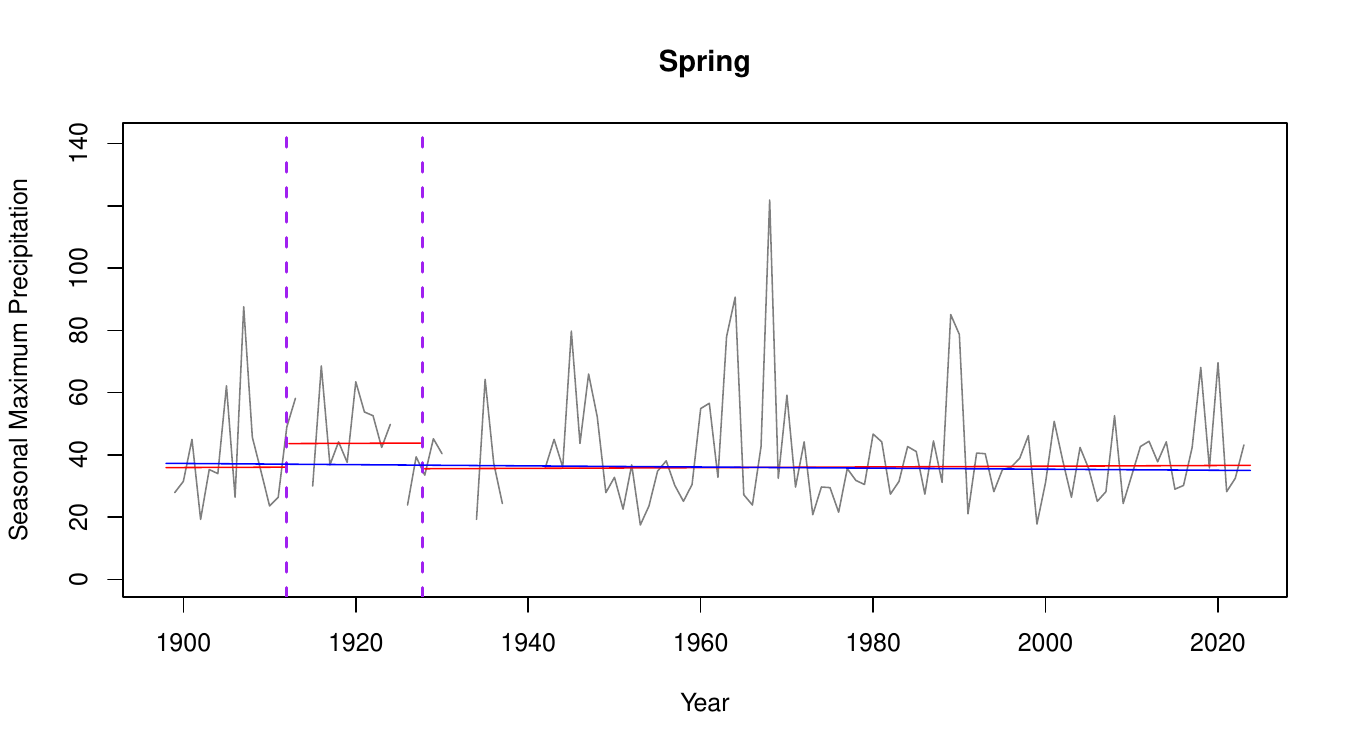}
           \hspace{-3mm}\includegraphics[width=0.54\linewidth]{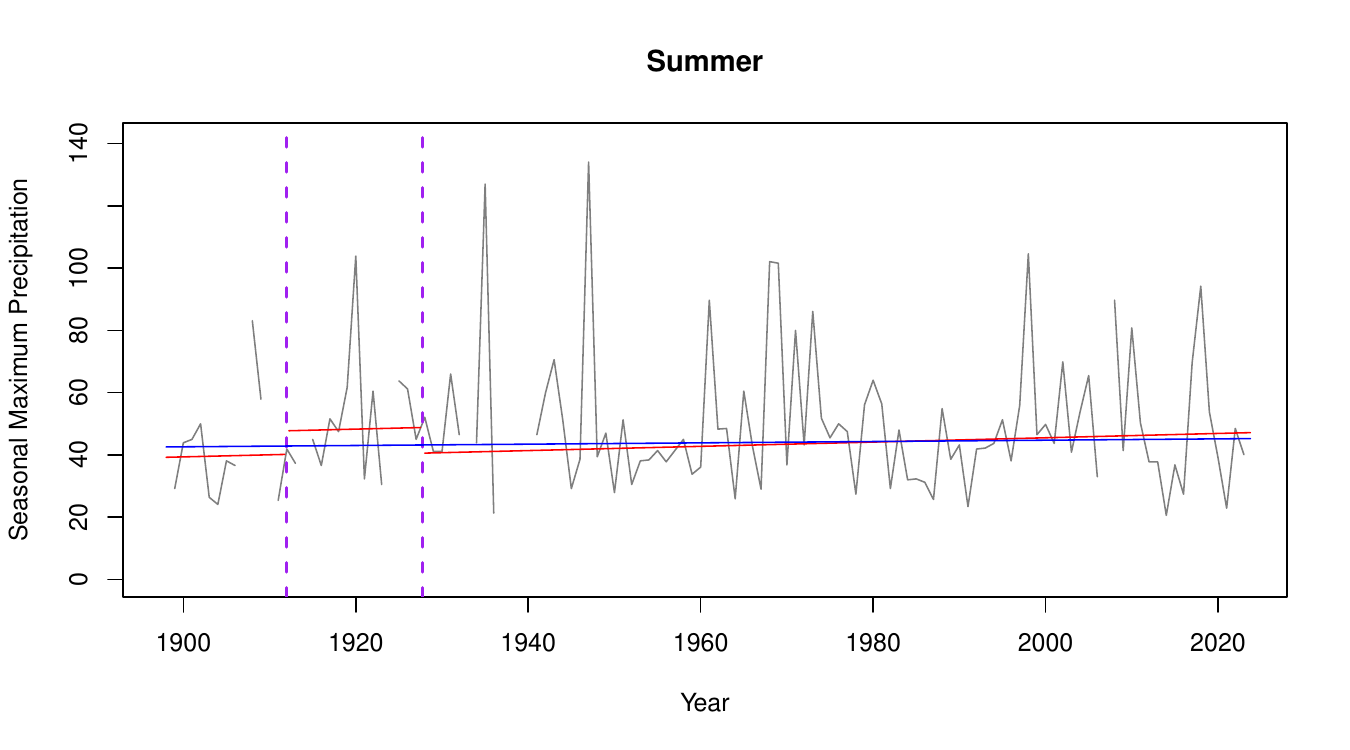}}\vspace{-3mm}
\centerline{\hspace{1mm}\includegraphics[width=0.54\linewidth]{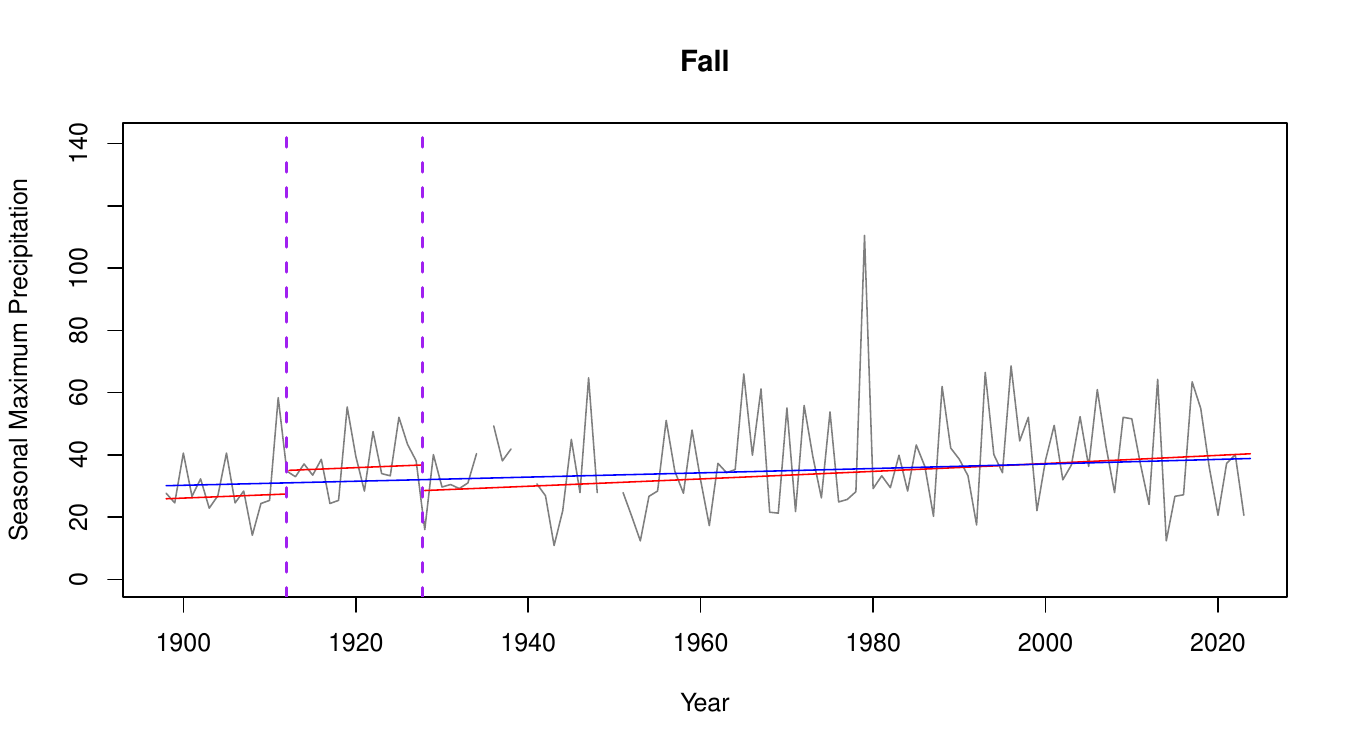}
           \hspace{-3mm}\includegraphics[width=0.54\linewidth]{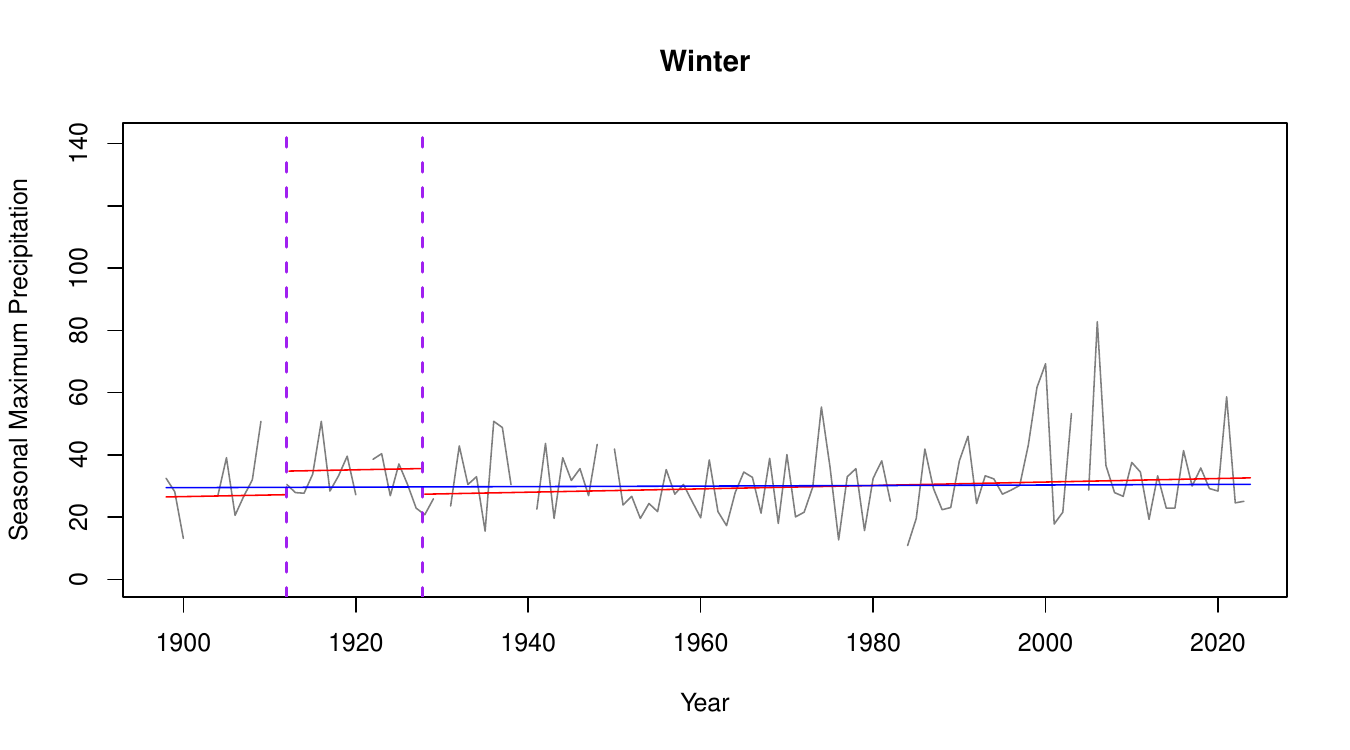}}
\vspace{-3mm}
\caption{Estimated GEV seasonal trend lines for the seasonal maximum precipitation in Circleville, OH (red solid line, estimated trend with GA changepoints considered; blue solid line, estimated trend with changepoints ignored; purple dashed lines, GA changepoint times)}
\label{f:Strends_Circleville}
\end{figure}

Now, we assess the fitness of our GEV model with GA-estimated changepoints. Figure~\ref{f:Strends_Circleville} compares the estimated seasonal trend lines for the Circleville station under two scenarios: with changepoints considered (red solid lines) and with changepoints ignored (blue solid lines). As verified in Section~\ref{s:cases_cpts}, the first changepoint in Spring 1912 aligns with documented changes in both station location and observer. Such movements of stations or relocations of rain gauges can induce discontinuities into precipitation records \citep{Groisman:Legates:1994}, while changes in observers may lead to subjective biases in precipitation measurements \citep{Daly:others:2007}. These alterations appear to have caused an abrupt increase in recorded precipitation starting in Spring 1912. The second changepoint, identified in Winter 1927, corresponds to a change in the OT schedule and is associated with a subsequent decrease in recorded precipitation. This pattern of `up-and-down' shifts also appears in other studies, such as the case shown in Figure~3b of \cite{Wang:others:2010}. Notably, incorporating changepoints results in stronger and more consistent positive trends across all seasons, with the most pronounced increase observed in fall. Overall, the trend lines that account for changepoints provide a better visual fit to the seasonal maximum precipitation series. This is further supported by the MDL criterion, with the MDL values of 3758.017 when changepoints are included and 3764.849 when they are not.

\begin{figure}[!ht]
\vspace{-4mm}
\centerline{\hspace{1mm}\includegraphics[width=0.54\linewidth]{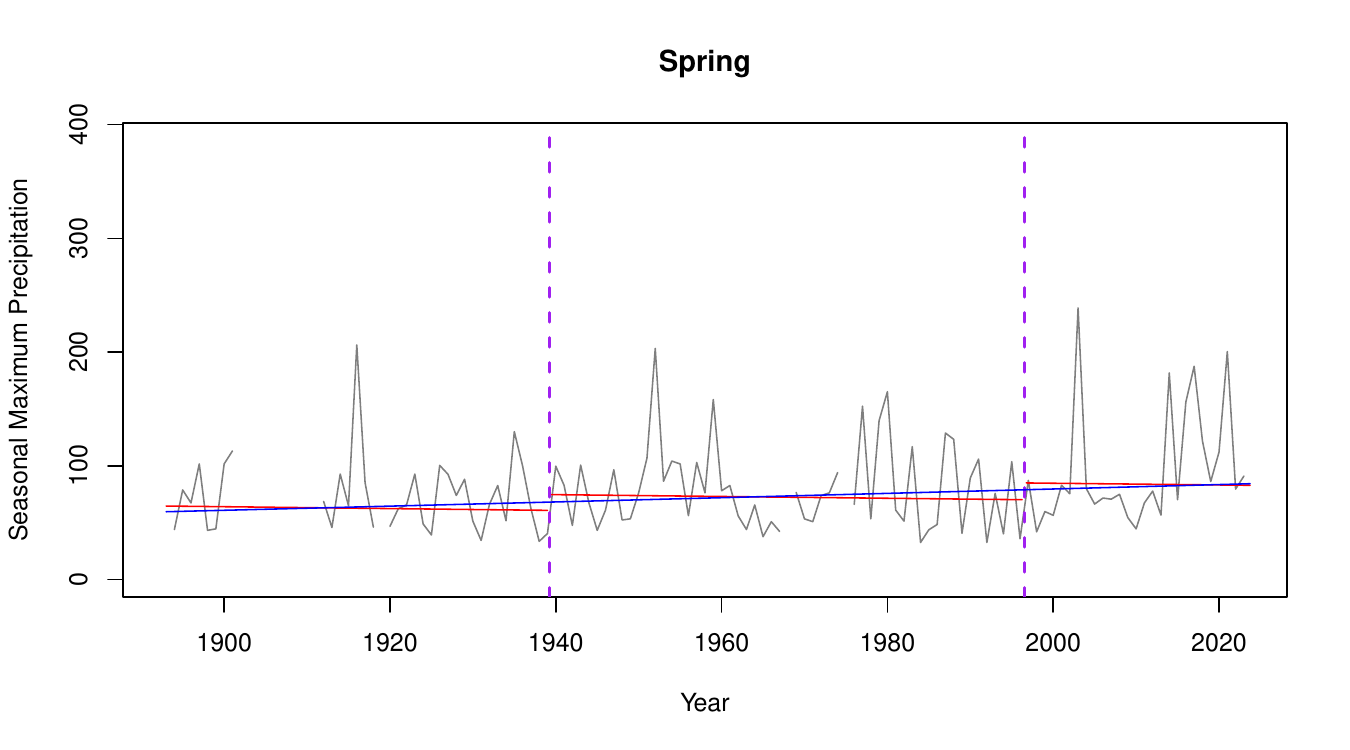}
           \hspace{-3mm}\includegraphics[width=0.54\linewidth]{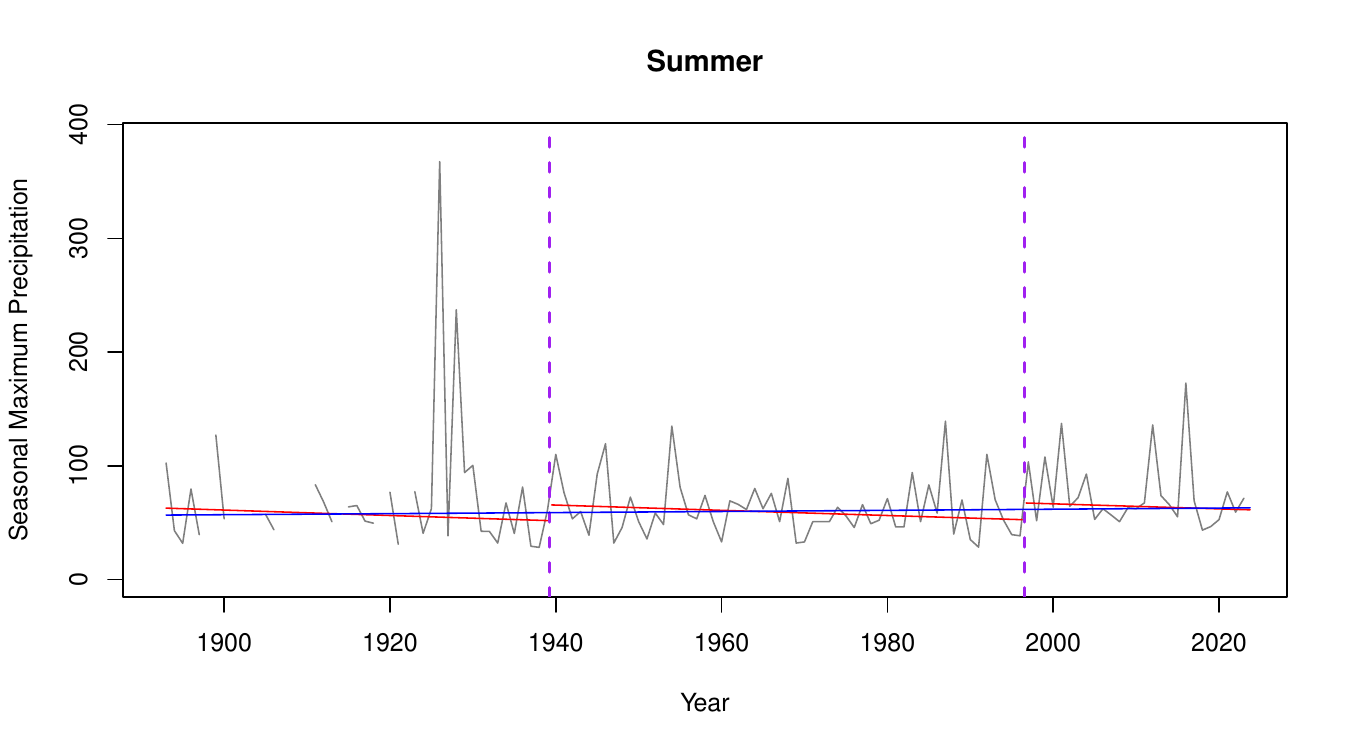}}\vspace{-3mm}
\centerline{\hspace{1mm}\includegraphics[width=0.54\linewidth]{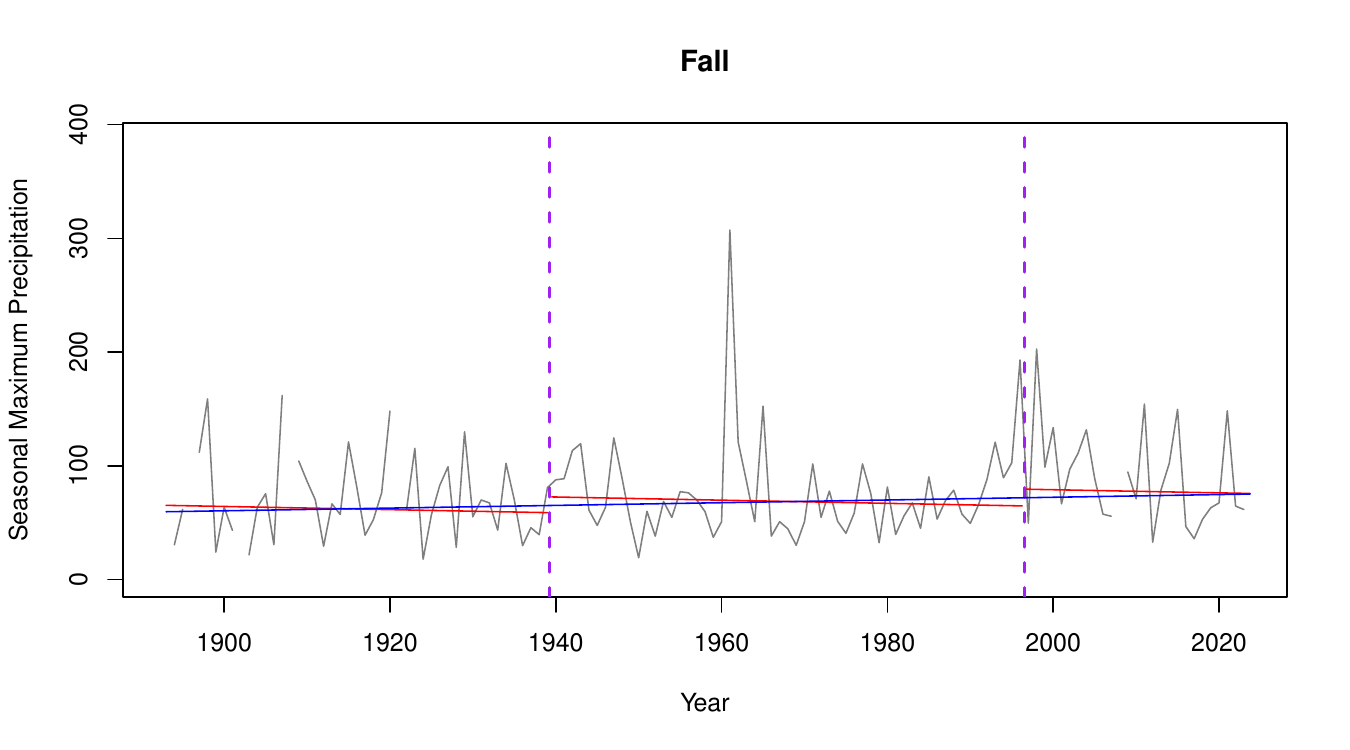}
           \hspace{-3mm}\includegraphics[width=0.54\linewidth]{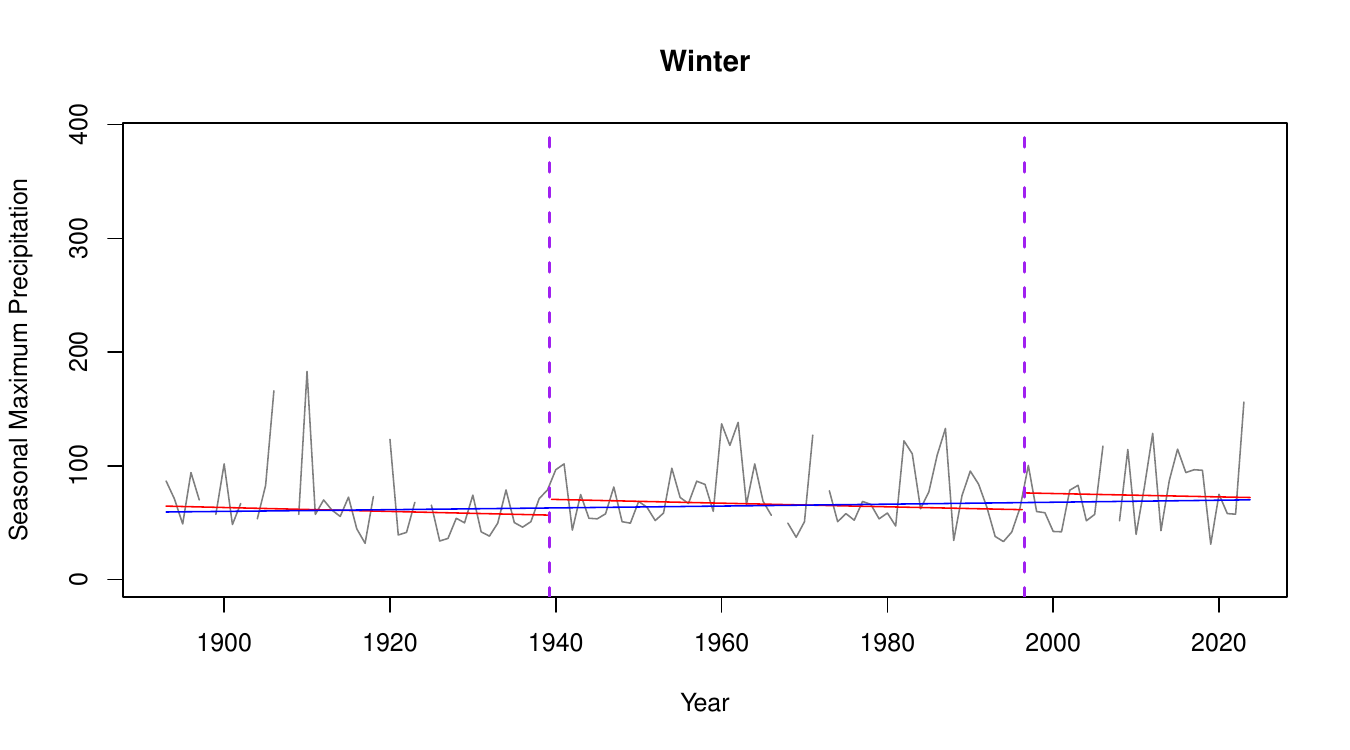}}
\vspace{-3mm}
\caption{Estimated GEV seasonal trend lines for the seasonal maximum precipitation in Donaldsonville, LA (red solid line, estimated trend with GA changepoints considered; blue solid line, estimated trend with changepoints ignored; purple dashed lines, GA changepoint times)}
\label{f:Strends_Donaldsonville}
\end{figure}

Similarly, Figure~\ref{f:Strends_Donaldsonville} demonstrates that the estimated seasonal trend lines for Donaldsonville also better fit the data when changepoints are accounted for. A series of changes in OT and instrument setup between September 1, 1918 and December 31, 1941 appears to cause an increase of precipitation measurements beginning in Summer 1939, the first detected changepoint at the Donaldsonville station. The installation and use of a Fischer-Porter gauge between July 1, 1995 and March 26, 2003 may contributed to another increase starting in Fall 1996, Donaldsonville's second changepoint. This pattern of `up-and-up' shifts is similar to findings such as the `down-and-down' pattern illustrated in Figure~3a of \cite{Wang:others:2010}. The MDL values are 4639.252 with GA-estimated changepoints and 4644.383 without them. These results reinforce the importance of incorporating changepoints in the trend analysis of seasonal maximum precipitation.

\begin{figure}[!ht]
\vspace{-6mm}
\centerline{\includegraphics[width=0.690\linewidth]{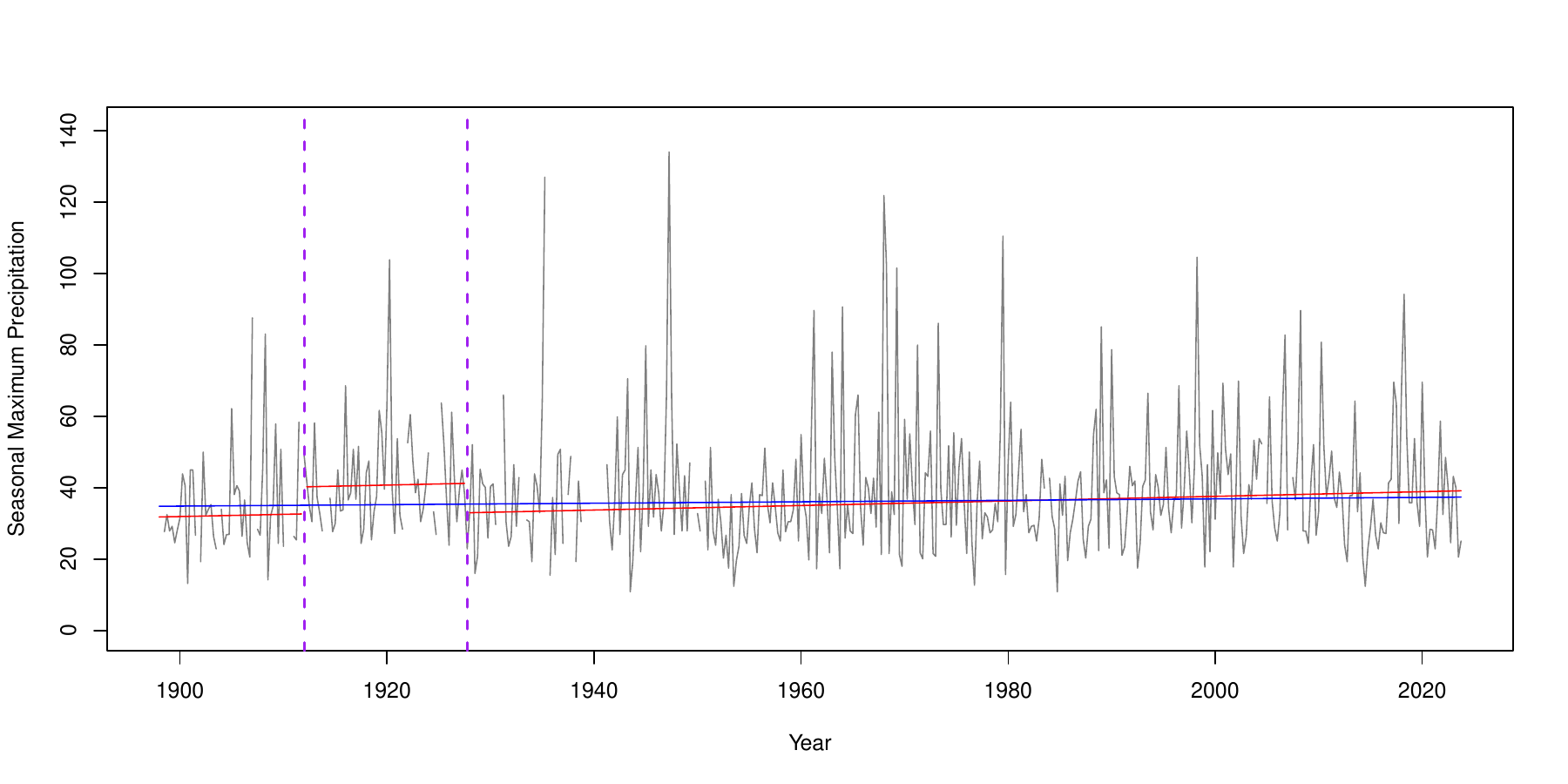}}\vspace{-8mm} 
\centerline{\includegraphics[width=0.692\linewidth]{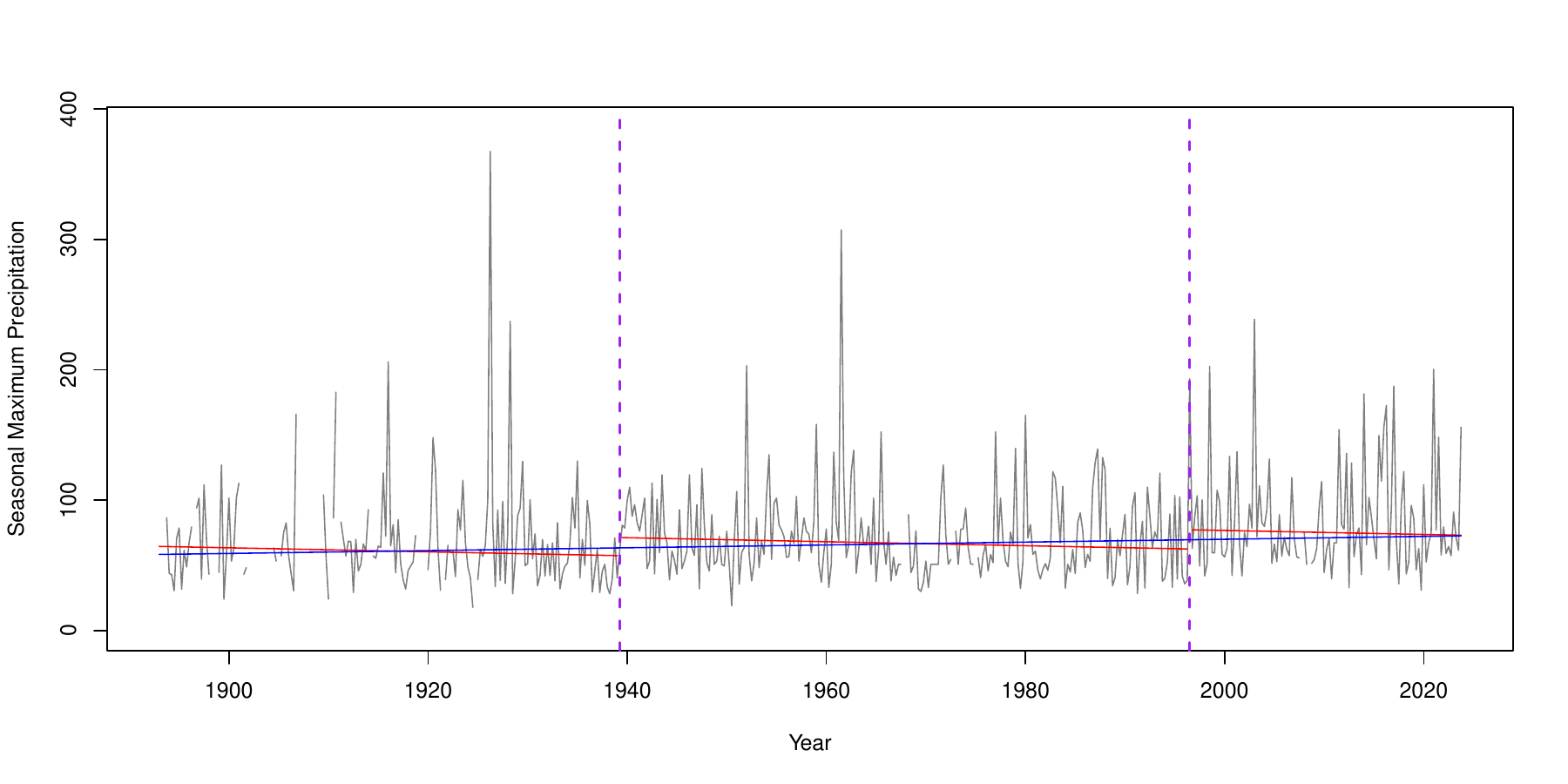}}              
\vspace{-3mm}
\caption{Estimated GEV long-term trend lines for the annual seasonal maximum average precipitation in Circleville, OH (top) and Donaldsonville, LA (bottom) (red solid line, estimated trend with GA changepoints considered; blue solid line, estimated trend with changepoints ignored; purple dashed lines, GA changepoint times)}
\label{f:Ltrends_cases}
\end{figure}

Next, we estimate the overall long-term linear trends as illustrated in Section~\ref{s:LT_rl_intrp}. For Circleville, the estimated long-term trend $\hat{\beta}_{\scriptscriptstyle\text{LTA}}$ in the annual average of seasonal extreme precipitation is 4.904 mm day$^{-1}$ century$^{-1}$ when GA-estimated changepoints are included, compared to 1.494 mm day$^{-1}$ century$^{-1}$ when changepoints are ignored. Considering their standard errors 2.170 and 1.539, we find a significant increase in the annual averaged seasonal extreme precipitation when changepoints are included but an insignificant increase when ignored. A similar pattern is observed for Donaldsonville. When changepoints are incorporated, the estimated long-term trend is --16.060 mm day$^{-1}$ century$^{-1}$ with a standard error of 6.026, indicating a significant decrease. In contrast, ignoring changepoints yields a significantly increasing trend of 9.725 mm day$^{-1}$ century$^{-1}$, with a standard error of 2.956. This contrast highlights how the presence or absence of changepoint adjustments can lead to opposite interpretations of long-term trends. Figure~\ref{f:Ltrends_cases} compares these estimated long-term trends for both stations. As supported by the MDL criterion, the trend lines that account for changepoints offer a better fit to the annual averages of seasonal extreme precipitation in both Circleville and Donaldsonville.

\subsection{Model diagnostics and seasonal return levels}\label{s:cases_rls}

\begin{figure}[!ht]
\vspace{-3mm}
\centerline{\includegraphics[width=0.50\linewidth]{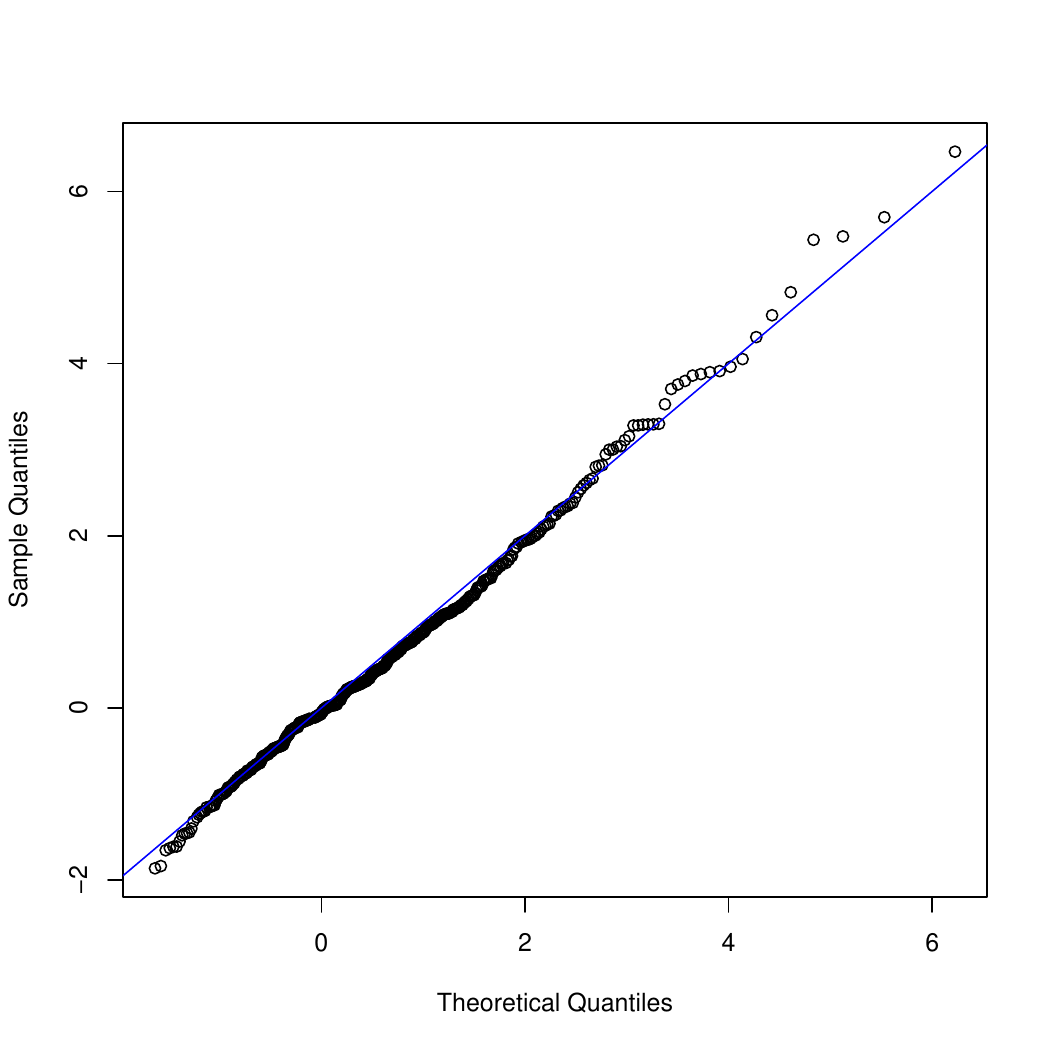}\hspace{-1mm} 
            \includegraphics[width=0.50\linewidth]{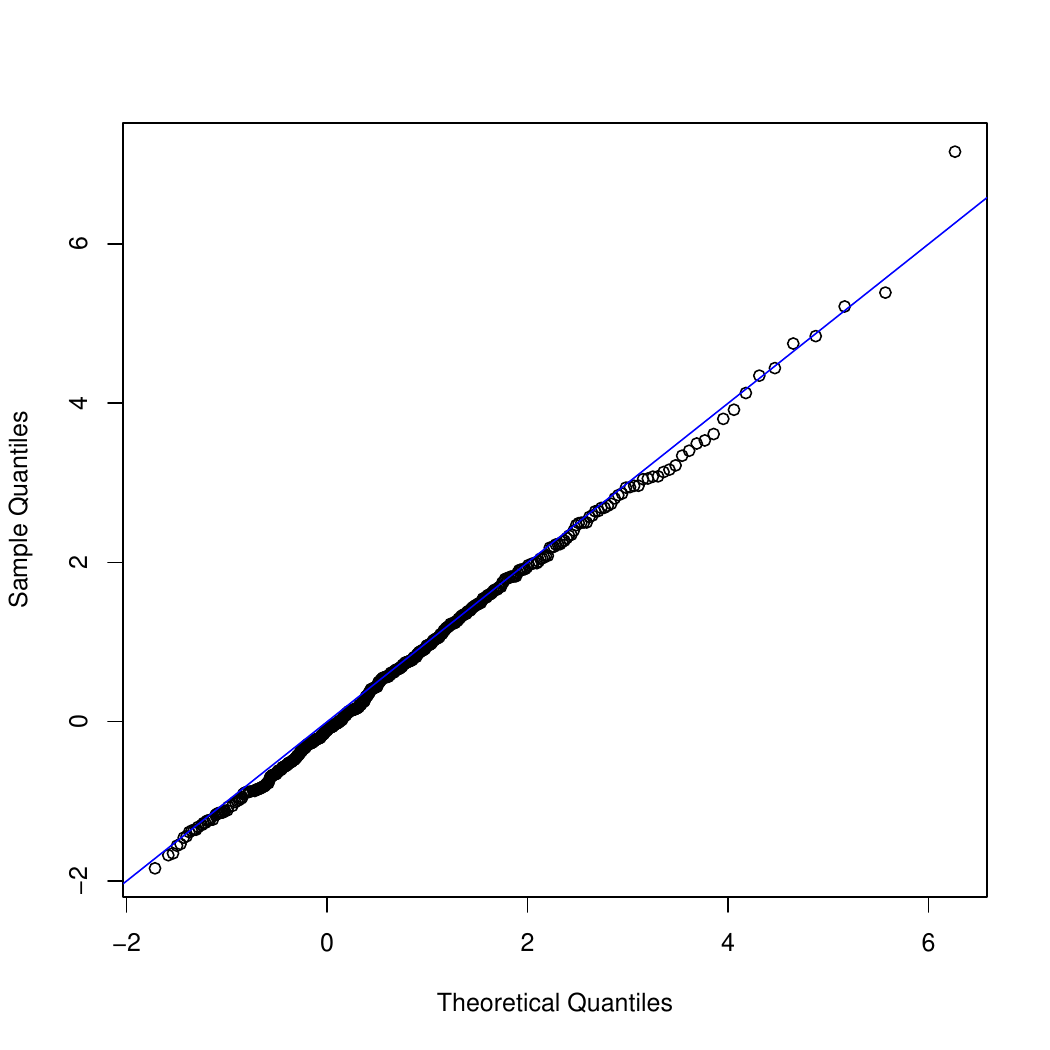}}             
\vspace{-3mm}
\caption{Gumbel-scaled Q-Q plots of the GEV models with GA changepoints for Circleville, OH (left) and Donaldsonville, LA (right)}
\label{f:QQplot_cases}
\end{figure}

To assess whether the fitted nonstationary GEV model with changepoints provides an appropriate fit for the seasonal maximum precipitation series, we examine the Gumbel-scaled Q-Q plot. This plot compares the sample quantiles of the seasonal maximum precipitation with the theoretical quantiles of the fitted GEV model after the transformation
\begin{equation*}
 \tilde{X}_t = \dfrac{1}{\hat{\xi}} \ln \left( 1 + \hat{\xi} \left( \dfrac{X_t - \hat{\mu}_t}{\hat{\sigma}_t} \right) \right).
\end{equation*}
The resulting transformed seasonal maximum series $\{ \tilde{X}_1, \ldots, \tilde{X}_n \}$ would then have a standard Gumbel distribution with the following cumulative distribution function
\begin{equation*}
 P(\tilde{X}_t \leq x) = \exp(-\exp(-x)).
\end{equation*}
The Q-Q plots in Figure~\ref{f:QQplot_cases} indicate that the fitted nonstationary GEV models with GA-estimated changepoints adequately capture the sample quantiles of the observed seasonal maximum precipitation at the Circleville and Donaldsonville stations.

Next, we estimate the 25- and 50-year seasonal return levels of the seasonal maximum precipitation for Circleville and Donaldsonville, as illustrated in Section~\ref{s:LT_rl_intrp}. Because climate dynamics often vary substantially across seasons in many US locations, return levels are estimated separately for each season, providing more relevant insights into seasonal extremes. To obtain accurate confidence intervals for return levels, we employed the bias-corrected and accelerated (BCa) bootstrap method \citep{Efron:1987, Givens:Hoeting:2013}, which adjusts for both bias and skewness in the bootstrap distribution. This approach makes the confidence intervals more reliable than conventional approaches such as the percentile bootstrap or the delta method. Given the inherent skewness of the GEV distribution, and consequently of return level estimates, the use of the BCa bootstrap is particularly well justified.

Table~\ref{t:rl_cases} summarizes the estimated 25- and 50-year return levels along with their associated 95\% BCa confidence intervals for both stations. These return levels represent the lowest seasonal maximum 24-hour precipitation expected to be exceeded, on average, once per season within the periods 2025--2049 (for the 25-year return level) and 2025--2074 (for the 50-year return level). Seasonal precipitation dynamics vary by station, with Circleville experiencing the highest maximum precipitation in summer and fall and the lowest in spring and winter, while Donaldsonville sees the highest in spring and the lowest in summer.

\vspace{1mm}
\begin{table}[!ht]
\caption{25- and 50-year maximum precipitation return levels with 95\% BCa bootstrap confidence intervals in parentheses (unit: mm day$^{-1}$)}
\label{t:rl_cases}
\vspace{1mm}
{\small
\begin{tabular*}{1.0\textwidth}{@{\extracolsep{\fill}}llcc@{\extracolsep{\fill}}}
\toprule
 Return level & Season & Circleville, OH & Donaldsonville, LA \\
\midrule
 \multirow{4}{*}{$r_{25}$}
    & Spring & \, 67.949 (54.729, \, 85.039) & 191.501 (139.187, 262.916) \\
    & Summer & \, 98.644 (76.142, 127.822) & 113.973 (\, 88.460, 151.925) \\
    & Fall   & \, 93.881 (74.184, 117.479) & 177.552 (132.995, 242.162) \\
    & Winter & \, 65.894 (50.387, \, 86.884) & 169.664 (128.066, 220.622) \\
\midrule
 \multirow{4}{*}{$r_{50}$}
    & Spring & \, 76.147 (58.691, 100.514) & 234.364 (158.464, 345.191) \\
    & Summer &   115.327 (84.328, 158.958) & 128.239 (\, 94.054, 183.921) \\
    & Fall   &   116.031 (86.823, 153.949) & 211.952 (149.865, 308.085) \\
    & Winter & \, 77.982 (56.167, 112.358) & 211.703 (150.617, 295.434) \\
\bottomrule
\end{tabular*}
}
\end{table}

\section{US seasonal trends and return levels}\label{s:USresults}

We applied the GA changepoint method to the seasonal maximum precipitation at each of the 1,057 USHCN stations. The number and time locations of changepoints were estimated for each location by minimizing the penalized log-likelihood function in (\ref{e:pllikl}), as demonstrated with the Circleville and Donaldsonville stations in Section~\ref{s:cases_cpts}. The method identified no changepoints at 634 stations (60.0\%), one changepoint at 228 stations (21.6\%), two changepoints at 132 stations (12.5\%), and three or more changepoints at 63 stations (6.0\%), with an average of 0.66 changepoints per station.

Next, we fitted the nonstationary GEV($\mu_{t},\sigma_{t},\xi$) model to the seasonal maximum precipitation series at each station, with $\mu_{t}$ and $\sigma_{t}$ specified in (\ref{e:gev_mu2}) and (\ref{e:gev_sigma2}), incorporating GA-estimated changepoints when detected. From the fitted models, we obtained GEV parameter estimates for seasonal maximum precipitation across all 1,057 stations. The seasonal trend estimates $\hat{\beta}_{1,s}$ were then spatially smoothed using a geostatistical method with the Mat\'ern covariance function in (\ref{e:cov_Matern}) for each season, as illustrated in Section~\ref{s:LT_rl_intrp}. This procedure yields predicted seasonal trends both for the 1,057 stations and for all unsampled locations. Based on preliminary analyses, we selected a smoothing parameter of $\nu = 1.01$ for spring, summer, and fall, and $\nu = 1.10$ for winter, to balance smoothness with the ability to capture local variation. The choice of a low-order $\nu$ limits spatial smoothing, which is appropriate given that extreme precipitation events are typically more localized than regional in nature.

\begin{figure}[!ht]
\vspace{-4mm}
\centerline{\includegraphics[width=0.54\linewidth]{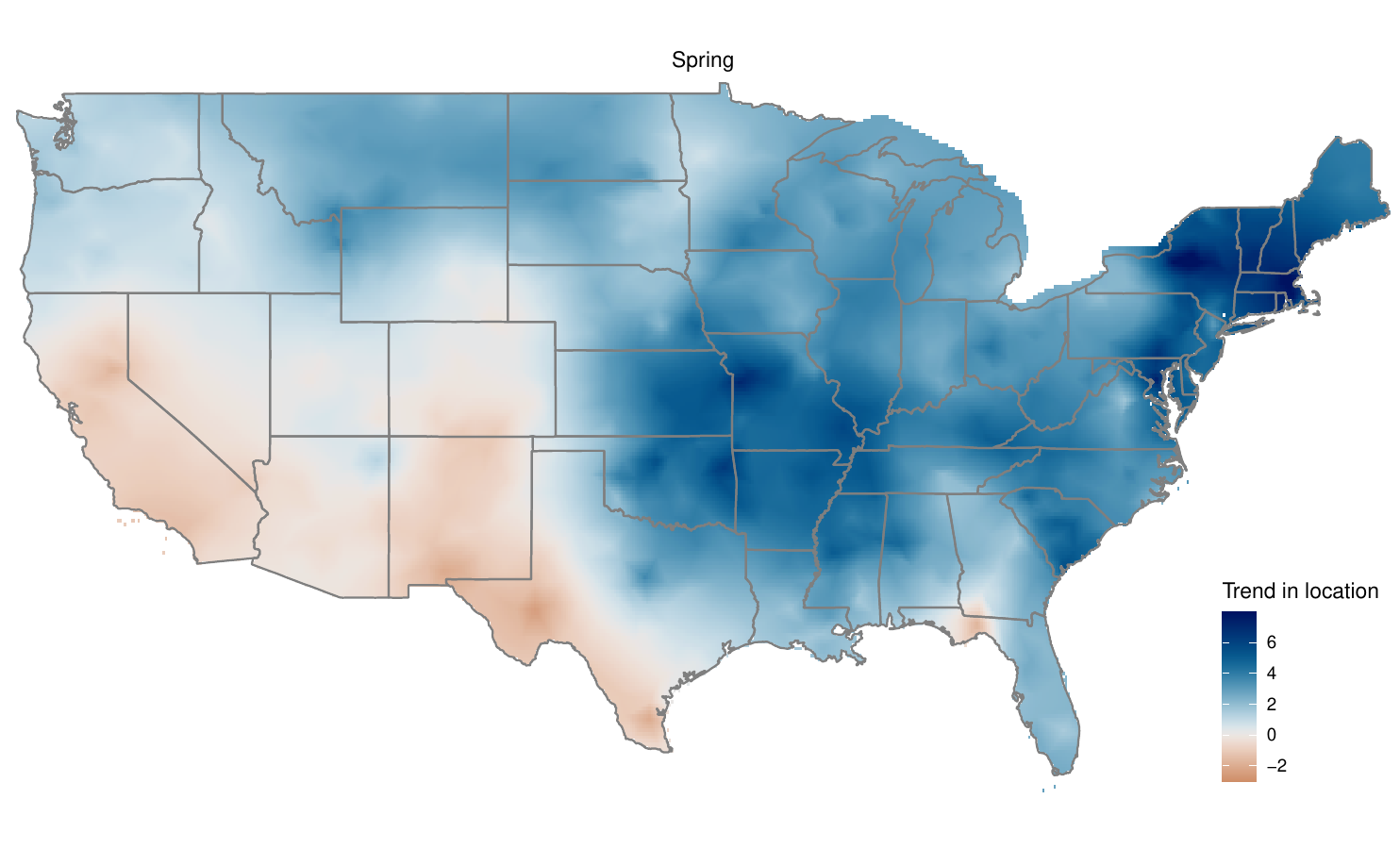}\hspace{-1mm}
            \includegraphics[width=0.54\linewidth]{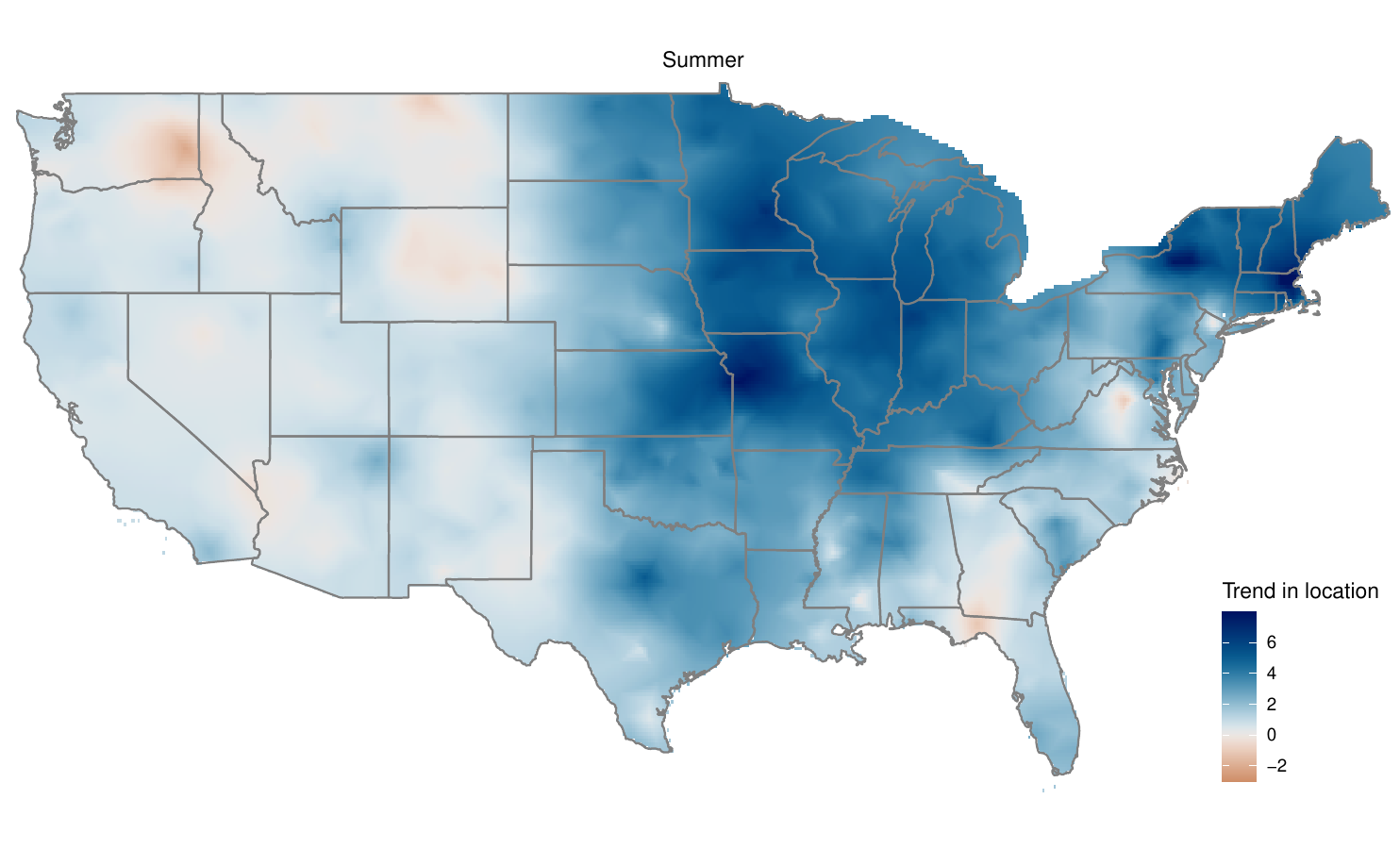}}\vspace{-3mm}
\centerline{\includegraphics[width=0.54\linewidth]{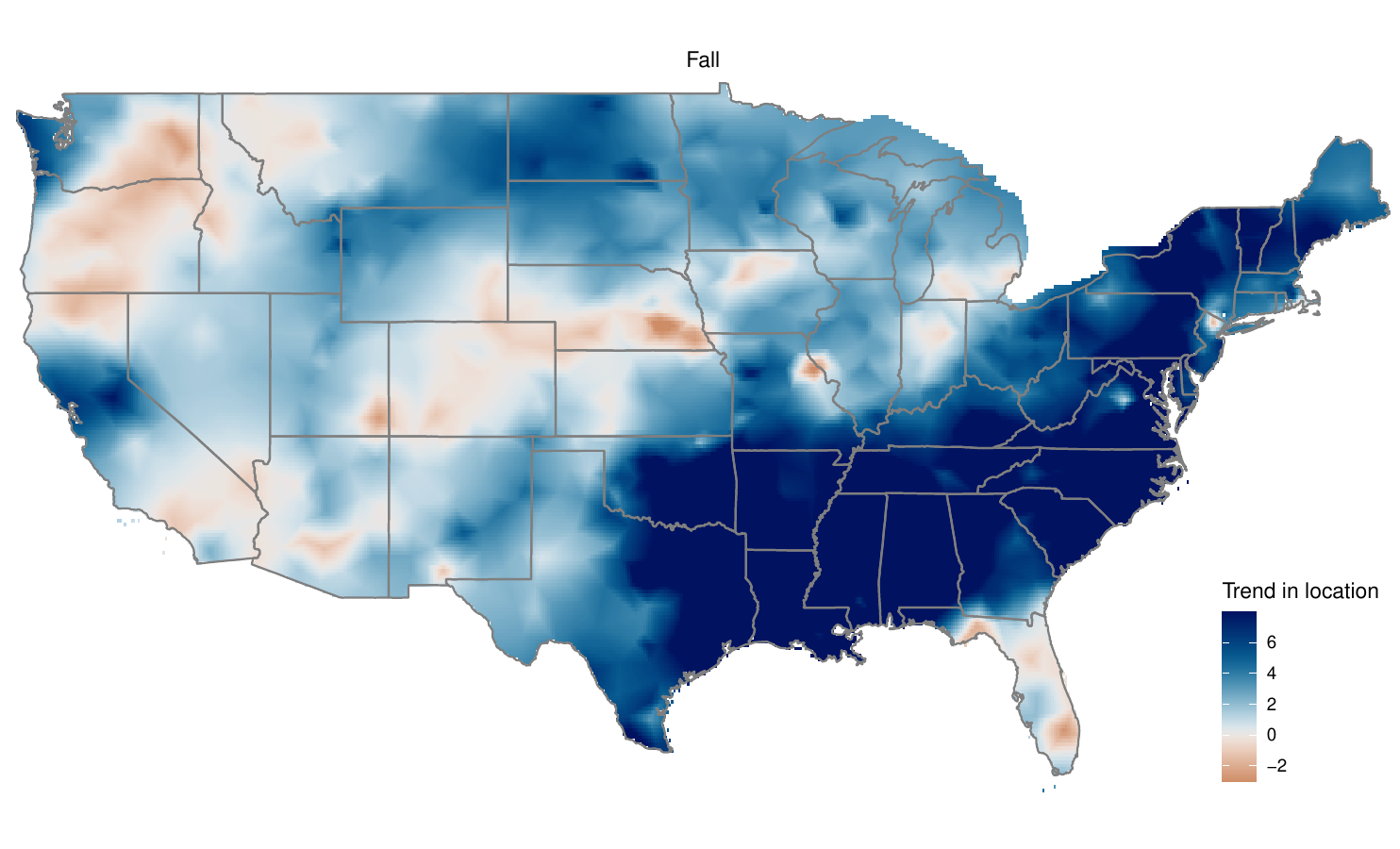}\hspace{-1mm}
            \includegraphics[width=0.54\linewidth]{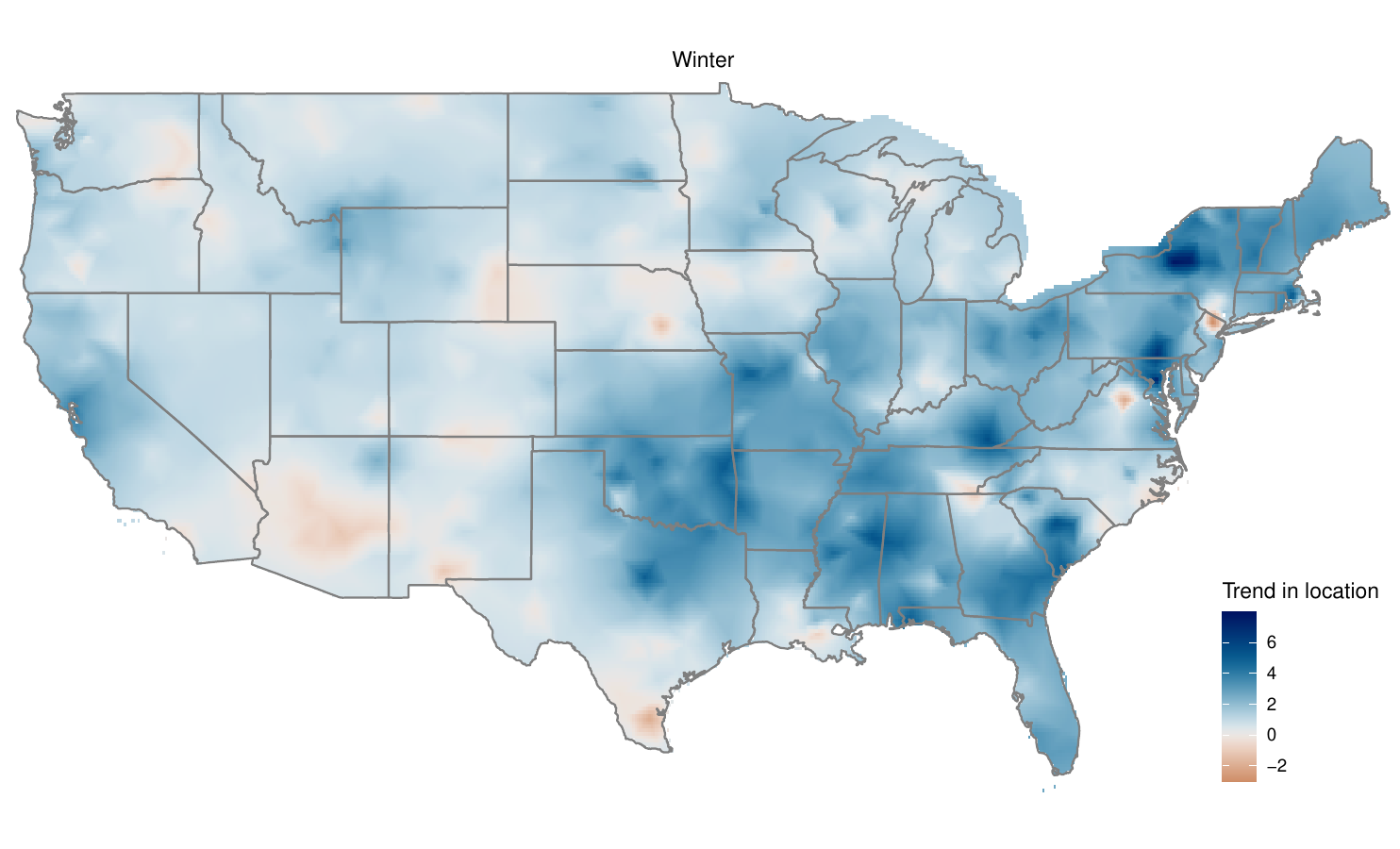}}
\vspace{-3mm}
\caption{Smoothed GEV seasonal trend estimates for US seasonal maximum precipitation: Spring (top left), Summer (top right), Fall (bottom left), and Winter (bottom right). Regions shaded in blue indicate increasing trends, while those in red indicate decreasing trends (unit: mm day$^{-1}$ century$^{-1}$).}
\label{f:STrends_US}
\end{figure}

We present the US seasonal trend estimation results in spatial maps. Figure~\ref{f:STrends_US} displays smoothed seasonal trend estimates from the GEV model, incorporating changepoints when detected. Overall, the maps indicate increasing maximum precipitation in the East and South, while also highlighting distinct regional and seasonal patterns. Specifically, spring maximum precipitation has increased in the Northeast and Central Plains but decreased in the Southwest, extending through western Texas. Summer trends show increasing maximum precipitation in the Northeast and Midwest but declining trends in the northern Pacific Northwest. In fall, increasing trends dominate from the Southeast to the East Coast, with rates exceeding 8 mm day$^{-1}$ century$^{-1}$ in most areas except Florida. Fall trends are also notably positive in the San Francisco Bay Area, western Washington, and Upper Midwest. In contrast, decreasing trends are observed from southeastern Washington through Oregon to northern California, as well as from southern California through Colorado to southern Nebraska. The winter trend map reveals localized increases in the Southeast, Northeast, and along the West Coast, while decreasing trends appear in Arizona and southern Texas. Overall, increasing trends are consistent across the South and East Coast in all seasons, particularly in fall. Meanwhile, decreasing trends are prominent in the Southwest during spring, especially in California, New Mexico, and southwestern Texas.

\begin{figure}[!ht]
\vspace{-4mm}
\centerline{\includegraphics[width=0.54\linewidth]{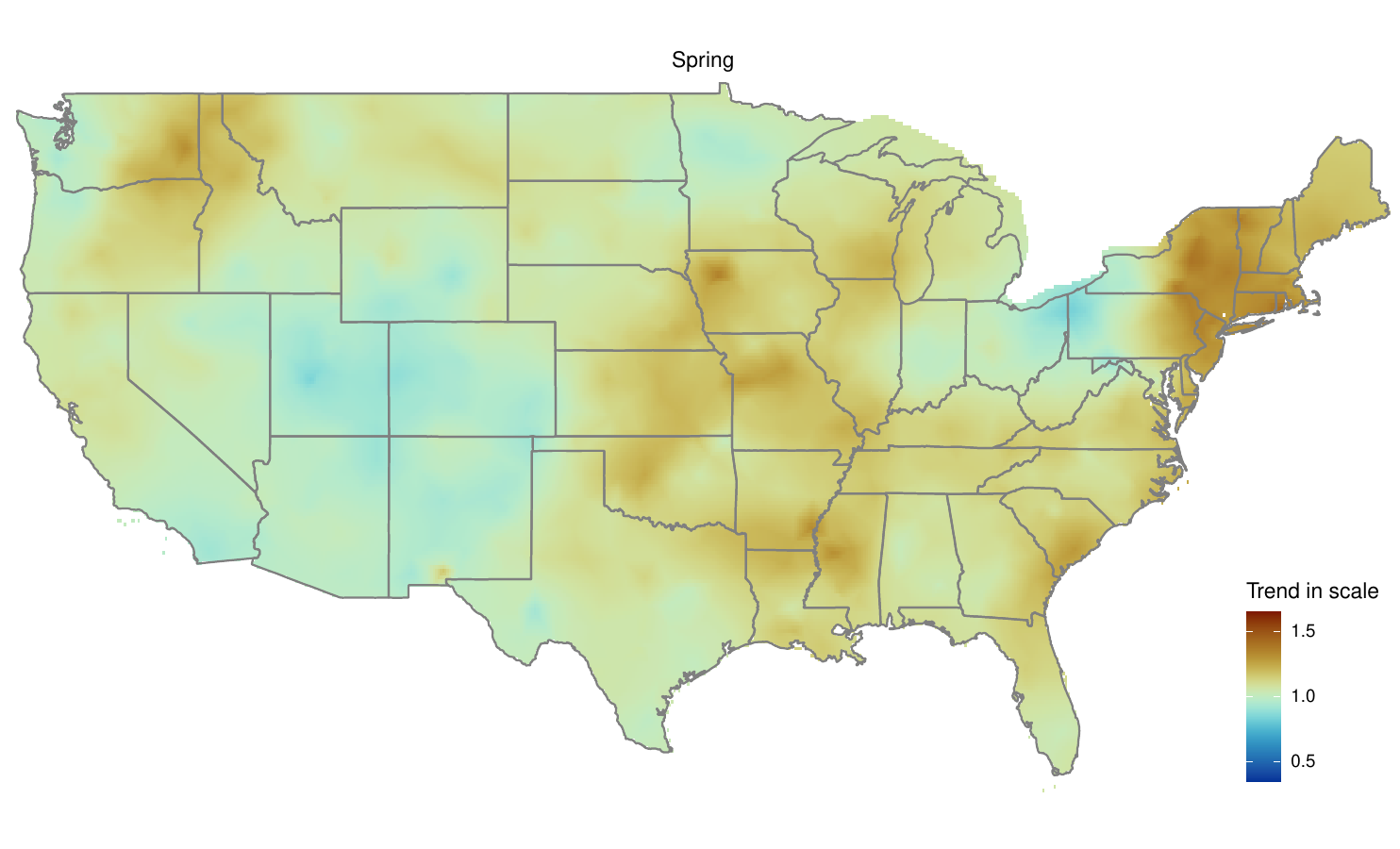}\hspace{-1mm}
            \includegraphics[width=0.54\linewidth]{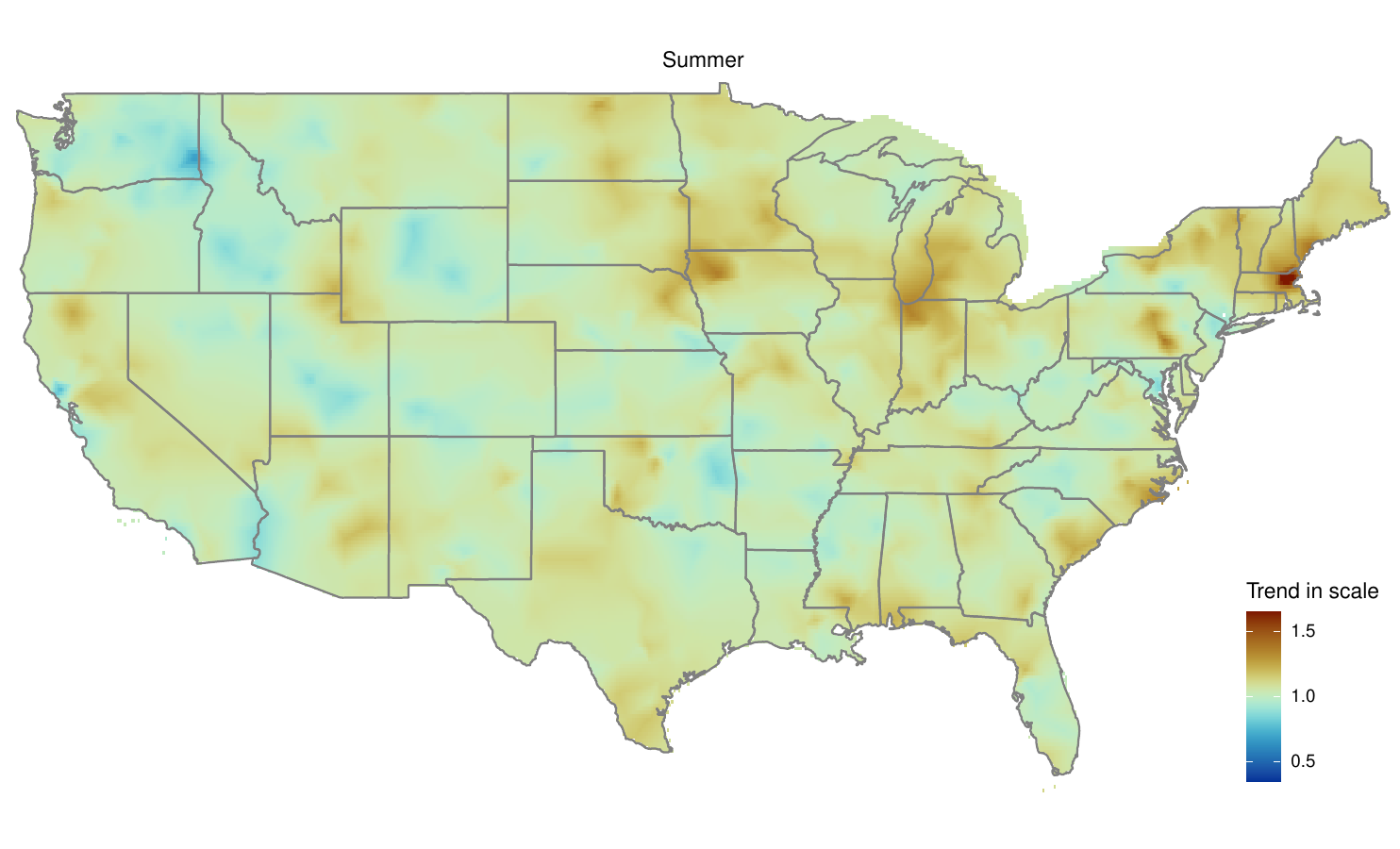}}\vspace{-3mm}
\centerline{\includegraphics[width=0.54\linewidth]{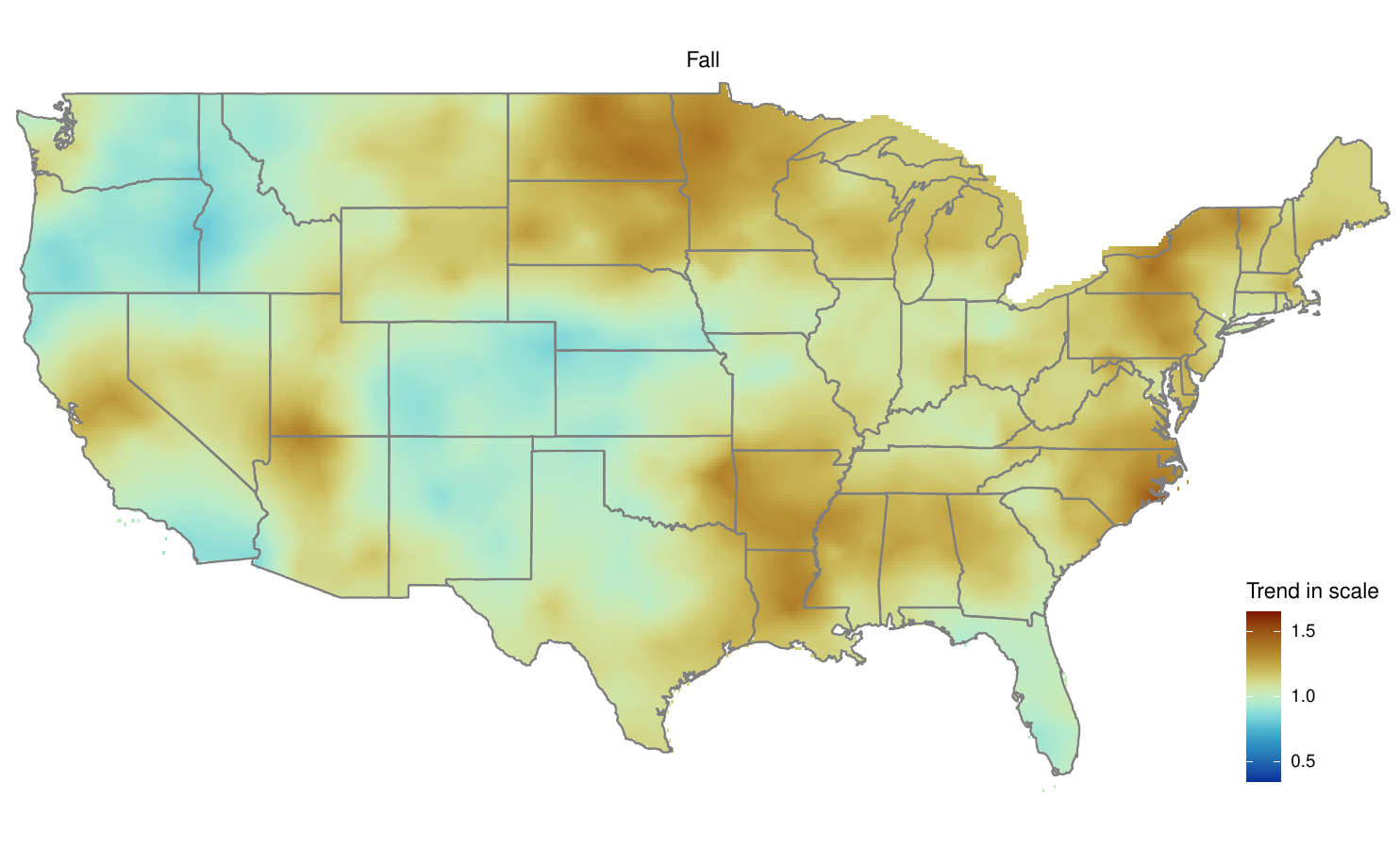}\hspace{-1mm}
            \includegraphics[width=0.54\linewidth]{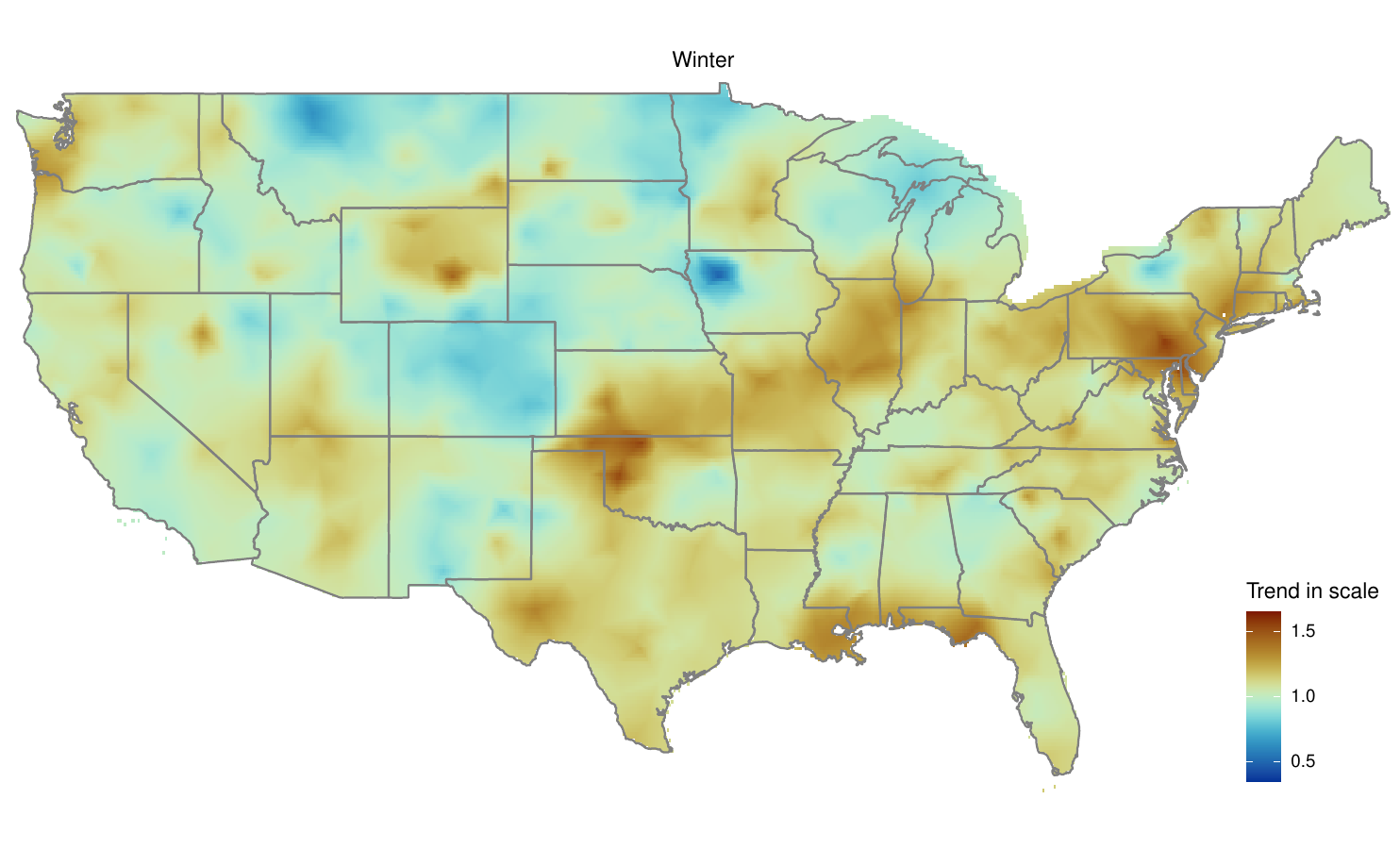}}
\vspace{-3mm}
\caption{Smoothed centennial changes in the seasonal standard deviation of US seasonal maximum precipitation: Spring (top left), Summer (top right), Fall (bottom left), and Winter (bottom right). Values indicate the multiplicative change in the standard deviation over a century, with red representing an increase and blue indicating a decrease.}
\label{f:SLambda_US}
\end{figure}

Another important aspect of seasonal maximum precipitation is the change in seasonal variance. Figure~\ref{f:SLambda_US} presents smoothed maps of the estimated $e^{\lambda_{1,s}}$ in (\ref{e:lambda}), which represents the multiplicative change in the standard deviation of seasonal maximum precipitation per century for each season $s=1,\ldots,T$. As explained in Section~\ref{s:gev}, values of $e^{\lambda_{1,s}}$ greater than one indicate increased variability, while values less than one indicate decreased variability over the past century. We selected smoothing parameters $\nu = 1.01$, $1.30$, $1.10$, and $1.10$ for spring, summer, fall, and winter, respectively, to balance spatial smoothness with local detail. The resulting maps reveal distinct seasonal patterns. In spring, variability has increased in the Northeast, along the East Coast, across the Central Plains, and in the eastern Washington--northern Idaho region, while it has decreased in Utah, Colorado, New Mexico, and along the Ohio--Pennsylvania border. Summer exhibits more localized changes, with pronounced increases along the East Coast and in northwestern Indiana, and decreases in Washington, southwestern Idaho, central Utah, and central Wyoming. In fall, increases are more widespread, spanning the Southeast, Northeast, Upper Midwest, Arizona, and from the San Francisco Bay Area through central Nevada, while decreases occur in the Pacific Northwest, Colorado, New Mexico, and southern California. Winter variability shows increases in Texas, western Oklahoma, Louisiana, Kansas, Missouri, Illinois, and Pennsylvania, with notable decreases in Montana, northwestern Iowa, and Colorado.

\begin{figure}[!ht]
\centerline{\includegraphics[width=0.68\linewidth]{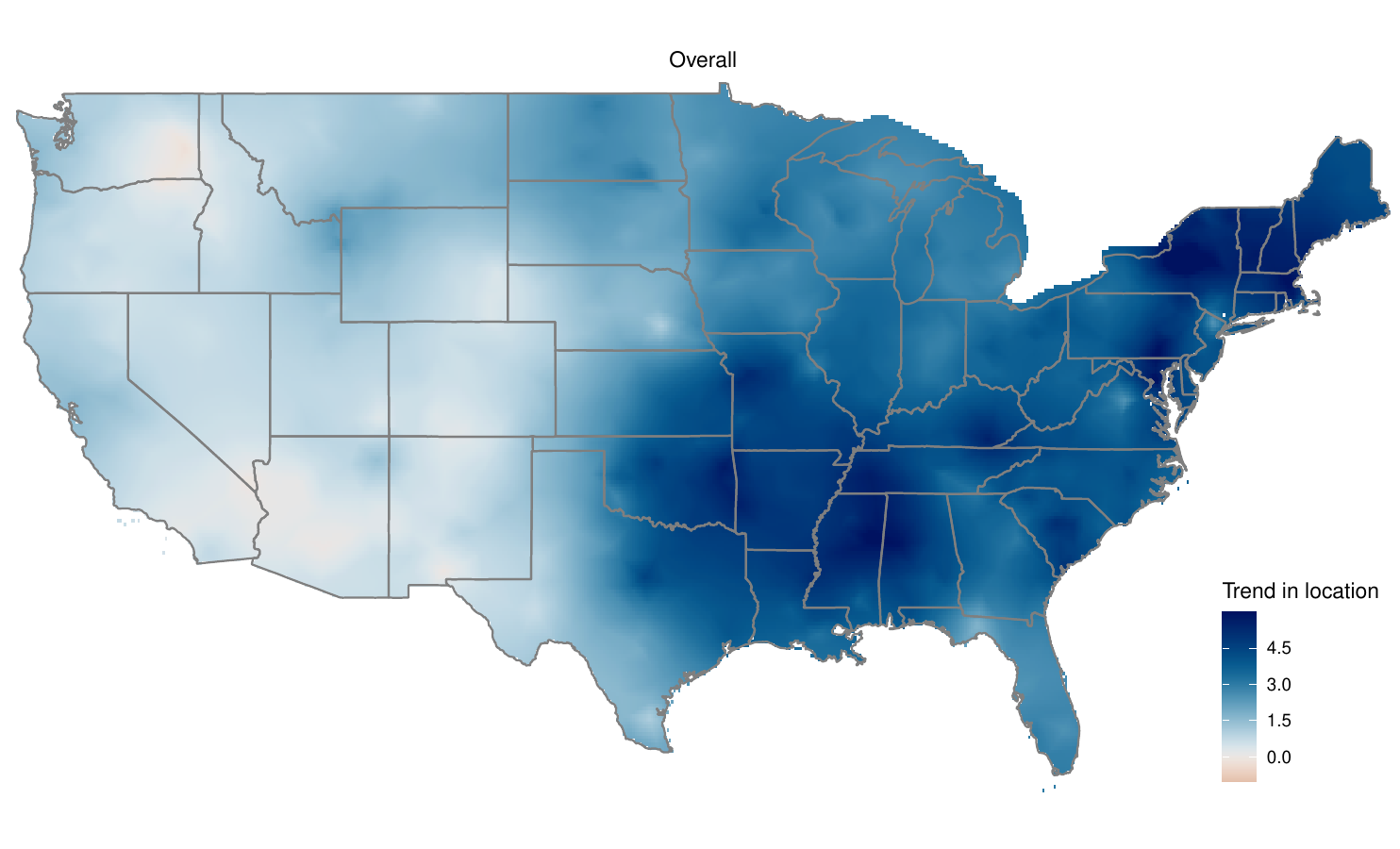}}
\vspace{-3mm}
\caption{Smoothed GEV long-term trend estimates for US annual seasonal maximum precipitation averages. Regions shaded in blue indicate increasing trends, while those in red indicate decreasing trends (unit: mm day$^{-1}$ century$^{-1}$).}
\label{f:All_Trends_US}
\end{figure}

Next, we examine long-term trends in seasonal maximum precipitation averages. Using the estimated seasonal trends, we computed long-term trend estimates $\hat{\beta}_{\scriptscriptstyle\text{LTA}}$ of annual seasonal maximum average precipitation across all 1,057 stations, as described in Section~\ref{s:LT_rl_intrp}. Figure~\ref{f:All_Trends_US} shows the spatial pattern of these long-term trends, smoothed with a smoothing parameter of $\nu = 1.01$ to retain local detail. This map indicates strong positive trends in the eastern US and weaker negative trends in the West. In particular, Louisiana, Arkansas, Missouri, Mississippi, and much of the Northeast show significant increases, whereas slight decreases are observed in the Southwest and the eastern Washington--northern Oregon region.

\begin{figure}[!ht]
\centerline{\includegraphics[width=0.54\linewidth]{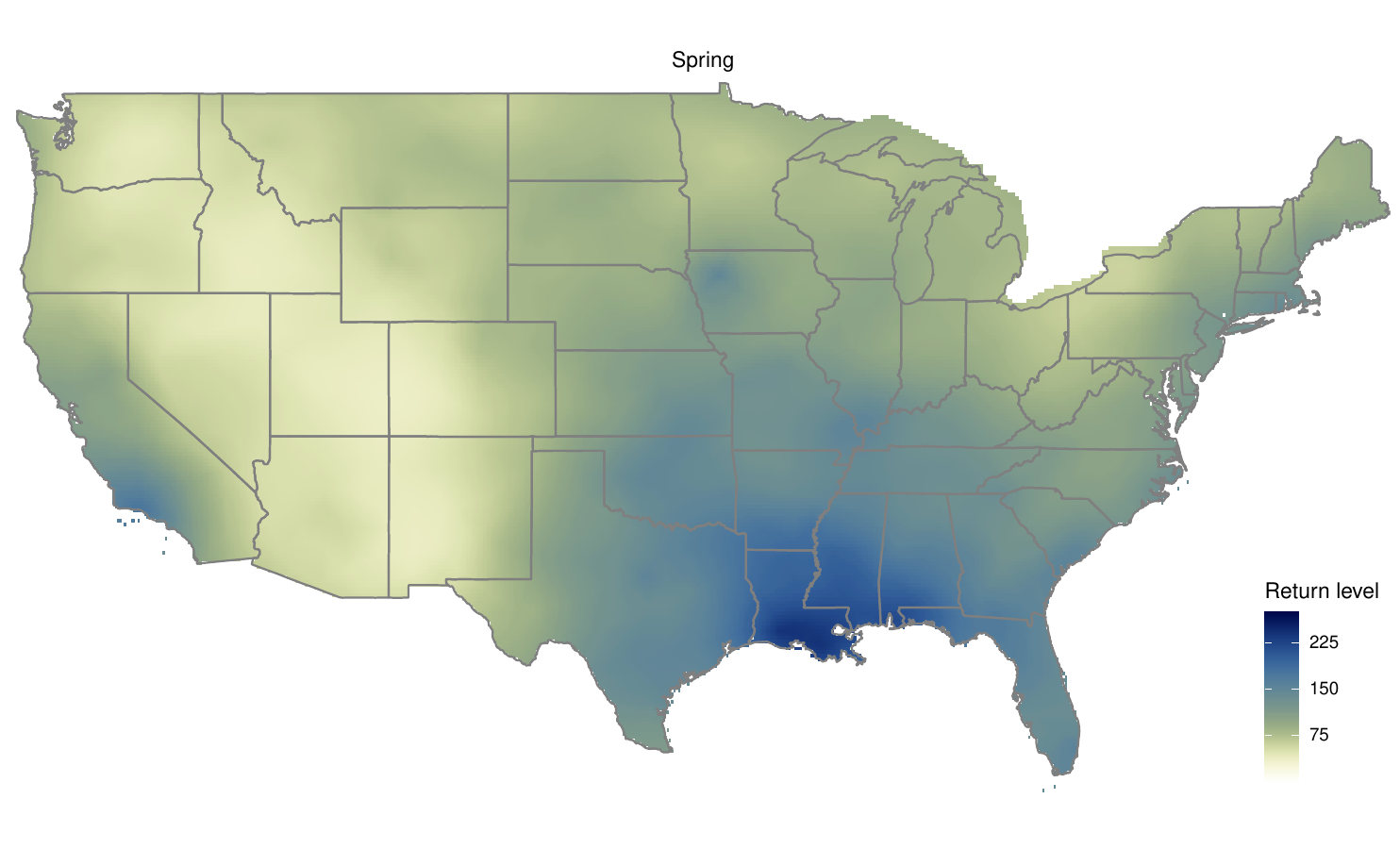}\hspace{-1mm}
            \includegraphics[width=0.54\linewidth]{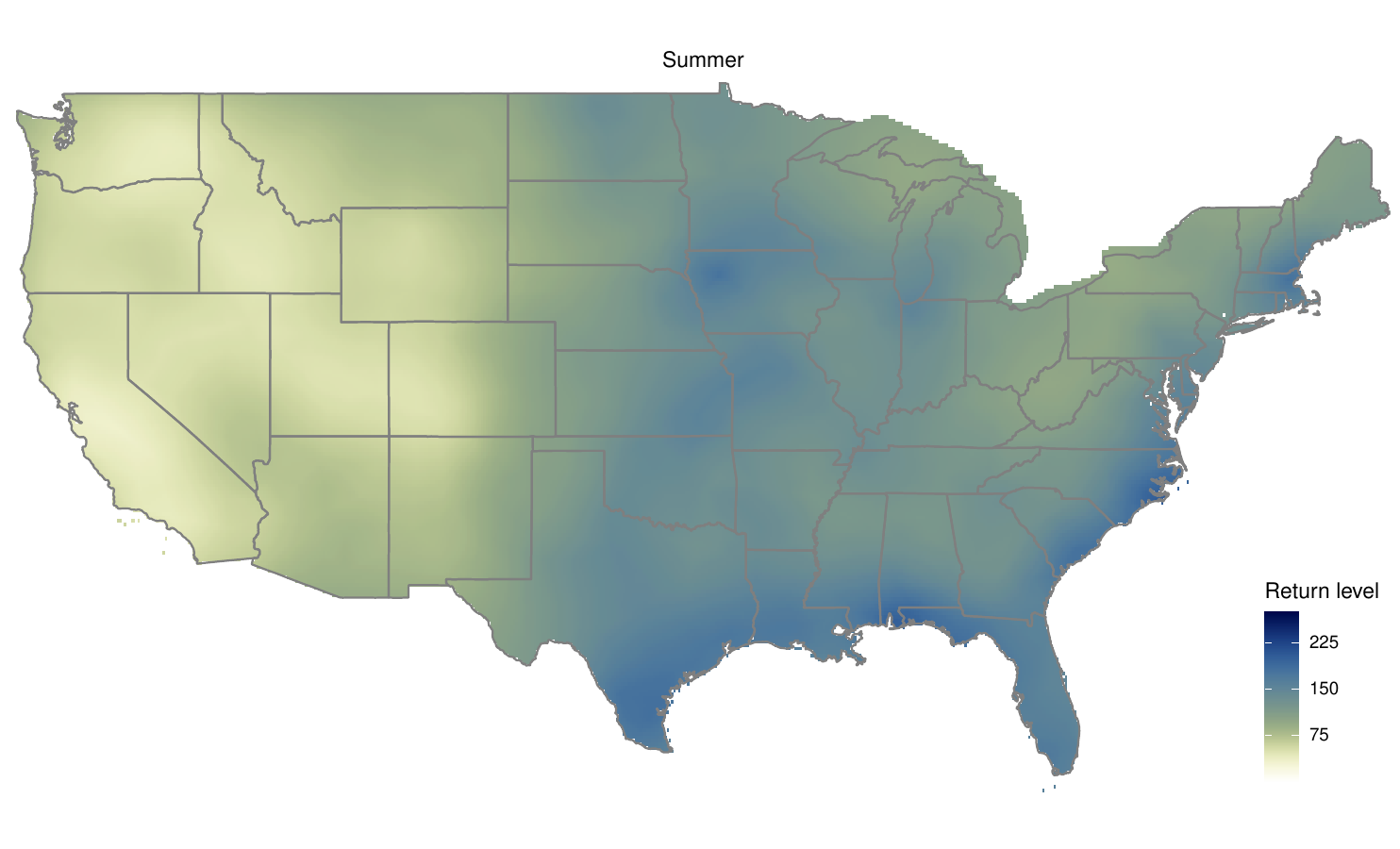}}\vspace{-3mm}
\centerline{\includegraphics[width=0.54\linewidth]{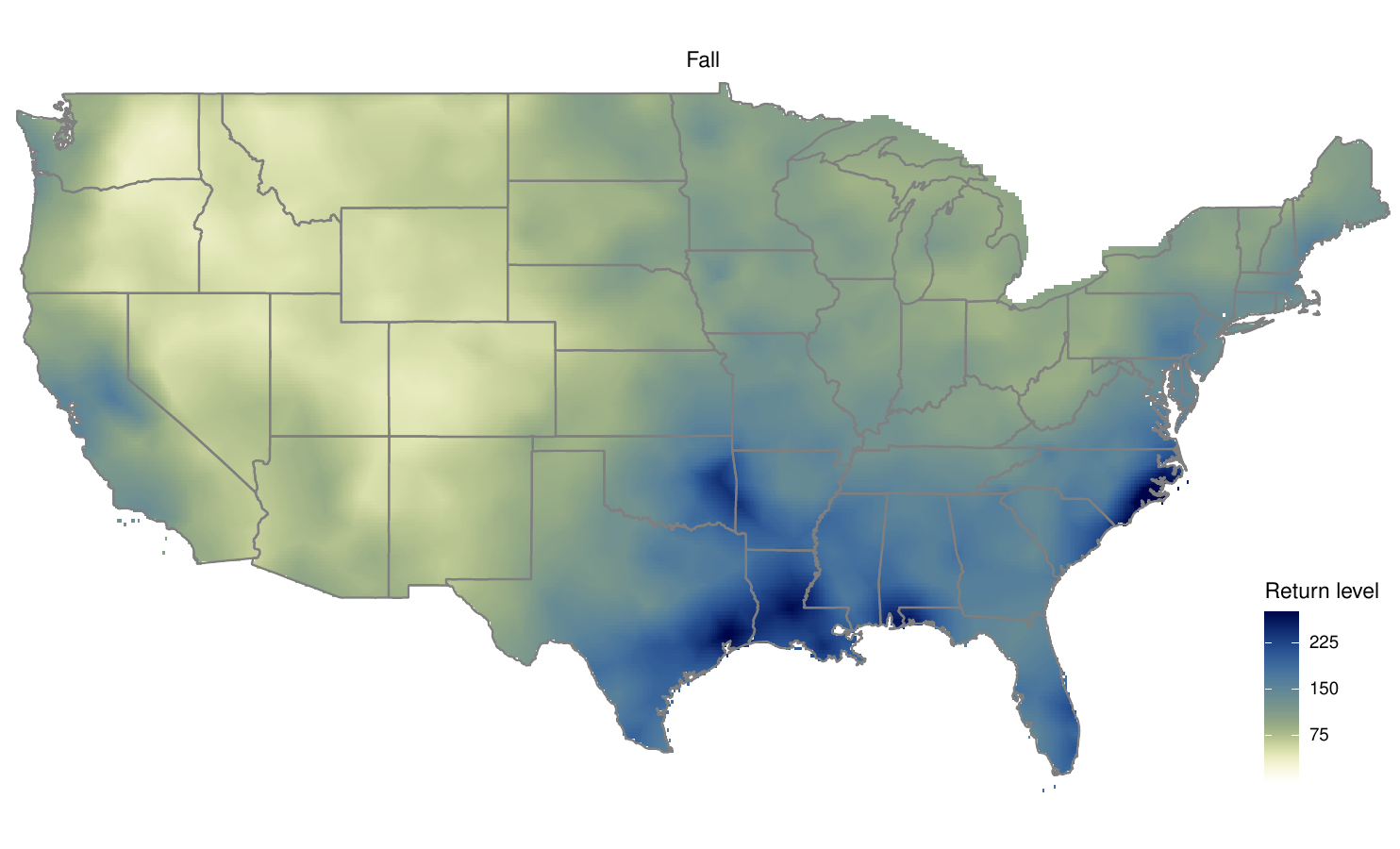}\hspace{-1mm}
            \includegraphics[width=0.54\linewidth]{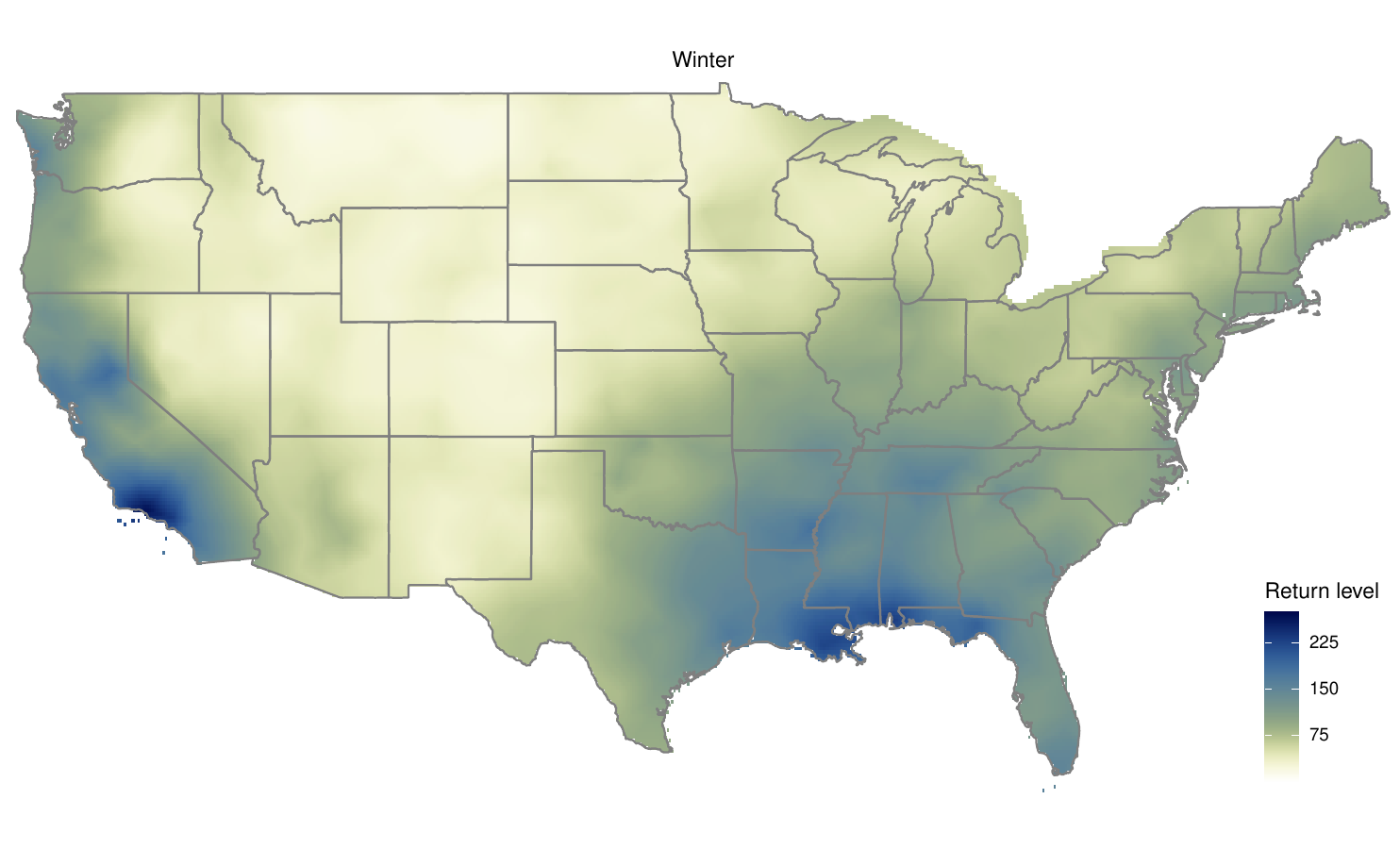}}
\vspace{-3mm}
\caption{Smoothed GEV seasonal 50-year return levels of US seasonal maximum precipitation: Spring (top left), Summer (top right), Fall (bottom left), and Winter (bottom right) (unit: mm day$^{-1}$)}
\label{f:Return50_US}
\end{figure}

We now present the seasonal return levels. Figure~\ref{f:Return50_US} presents the spatial patterns of smoothed 50-year seasonal maximum precipitation return levels, obtained using a smoothing parameter of $\nu = 1.11$ to maintain consistent smoothness across seasons while preserving local detail. Overall, eastern states are expected to experience higher seasonal maximum precipitation than western states, with much of the South anticipating particularly heavy rainfall. While broad spatial patterns are similar across seasons, notable regional differences emerge for different seasons. For example, the southern California coastline is projected to receive heavy rainfall in spring and winter, the Midwest may experience more extreme precipitation in summer, and eastern coastal states could see heavier rains in both summer and fall. In the South, high precipitation is expected in spring, fall, and winter, with fall showing the most intense rainfall. By contrast, winter is generally associated with lower maximum precipitation across most of the US, except along the California coastline, where heavier precipitation is anticipated.

\section{Discussion and conclusions}\label{s:conclusion}

\begin{figure}[!ht]
\vspace{-6mm}
\centerline{\includegraphics[width=0.825\linewidth]{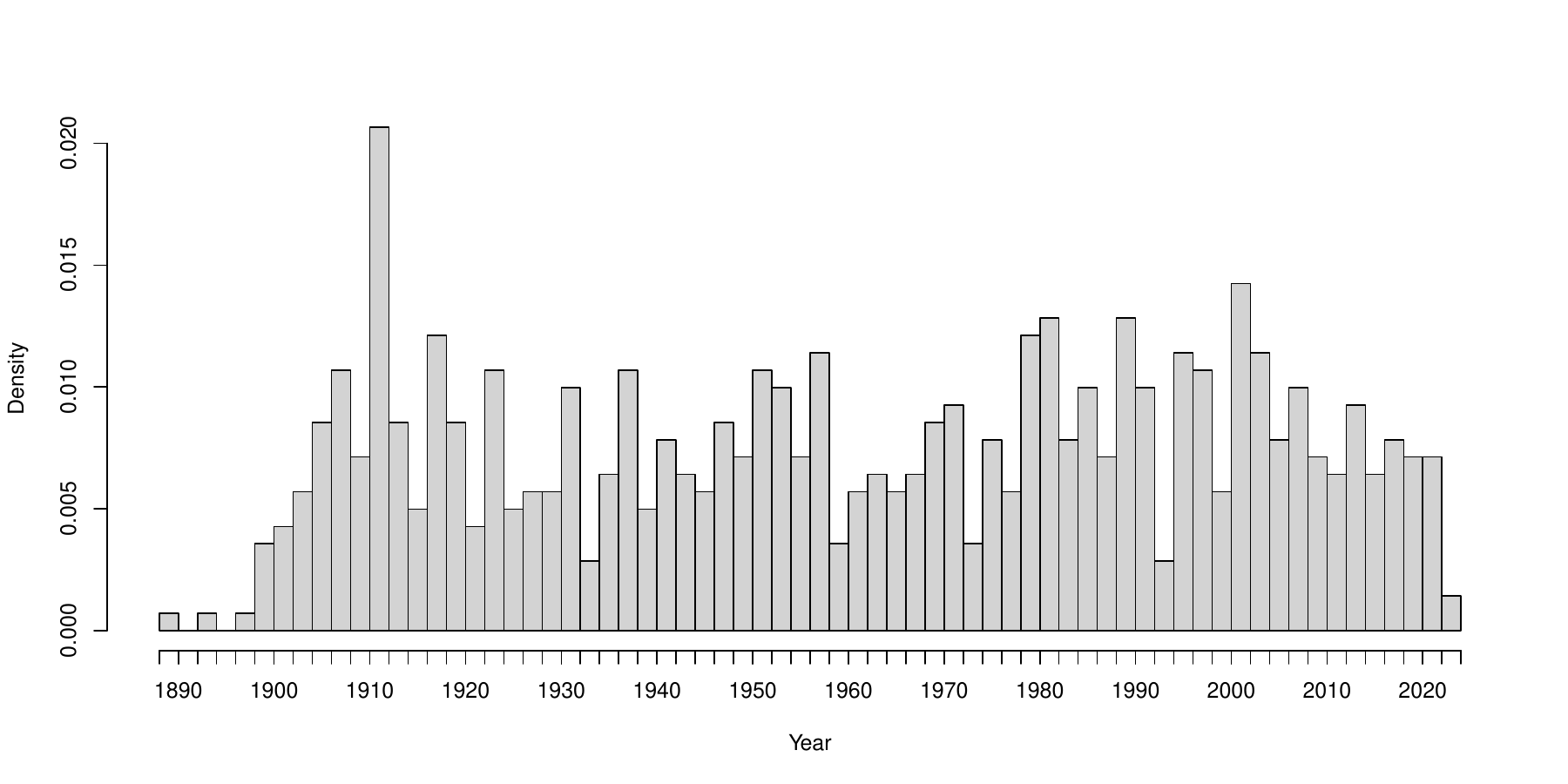}}
\vspace{-4mm}
\caption{Histogram of the GA estimated changepoint times for the selected 1,057 USHCN stations}
\label{f:Hist_cp_times}
\end{figure}

To explore potential causes of changepoints in extreme precipitation, we aggregate the GA-estimated changepoint times identified in our analysis. Figure~\ref{f:Hist_cp_times} presents a histogram of all changepoint times from the 423 stations with at least one changepoint detected. The most frequent changepoint occurrences are in the years 1910--1911, 1916--1917, 1922--1923, 1950--1958, 1978--1981, 1988--1991, and 2000--2003. While not all inhomogeneities resulting from changes in instrumentation or OT can be explicitly identified \citep{Applequist:others:2024}, some of these periods appear to coincide with known changepoint-inducing events. Previous research has noted that inhomogeneities arise due to observation time adjustments after 1950, as well as other factors such as instrument modifications and station relocations in the mid-1980s \citep[cf.][]{Menne:others:2009, Menne:others:2018}.

\begin{figure}[!ht]
\vspace{-4mm}
\centerline{\hspace{1mm}\includegraphics[width=0.45\linewidth]{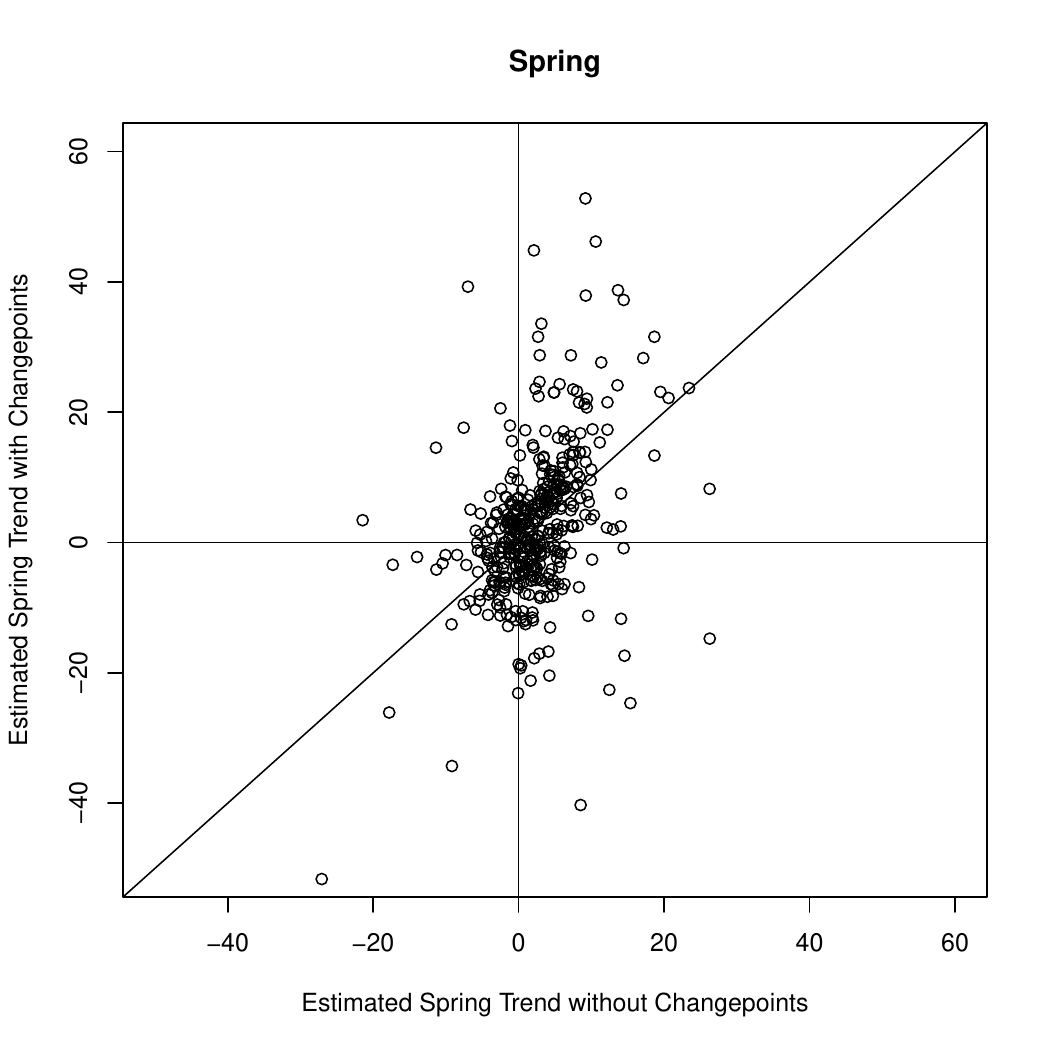}
           \hspace{-0mm}\includegraphics[width=0.45\linewidth]{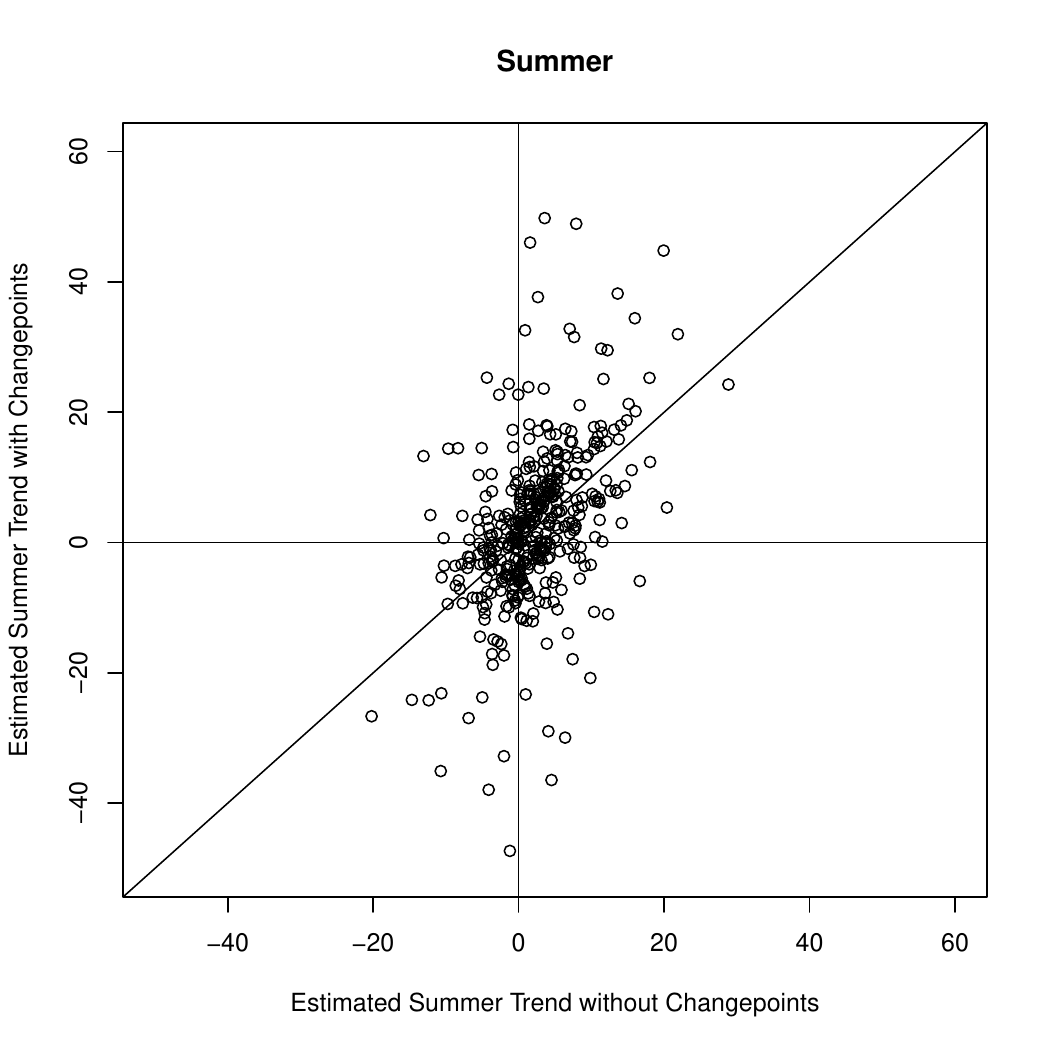}}\vspace{-2mm}
\centerline{\hspace{1mm}\includegraphics[width=0.45\linewidth]{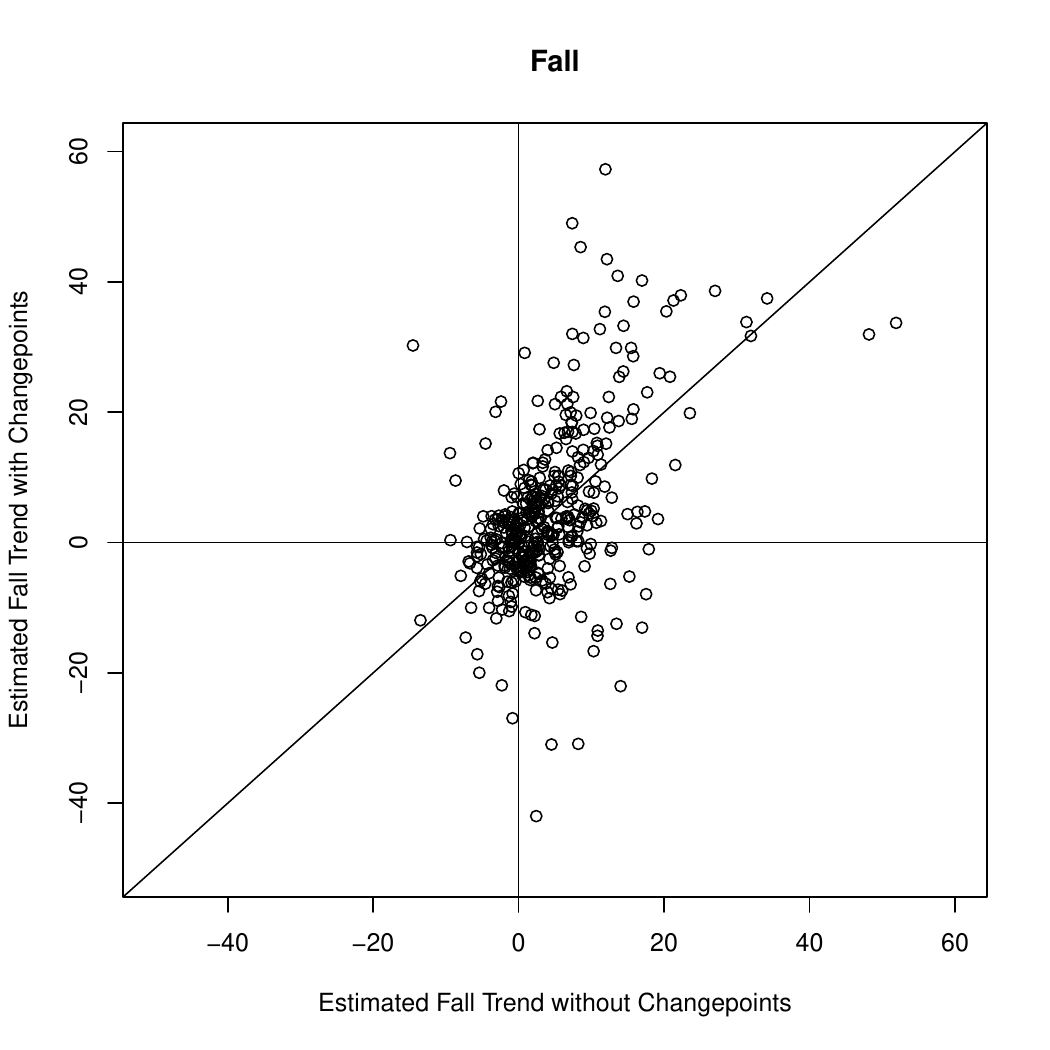}
           \hspace{-0mm}\includegraphics[width=0.45\linewidth]{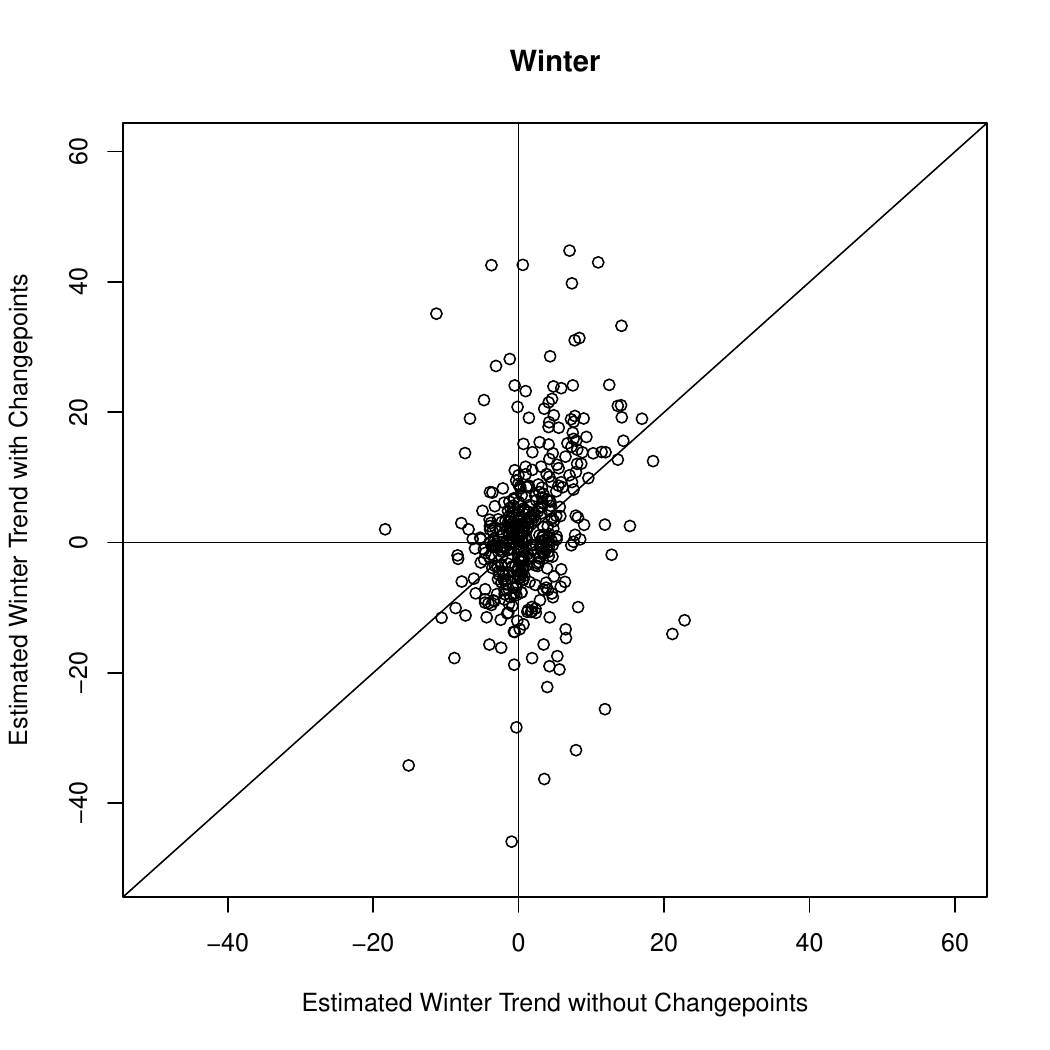}}
\vspace{-3mm}
\caption{Scatterplots of the estimated seasonal trends with GA-estimated changepoints considered against the changepoints ignored}
\label{f:Scatter_Strends}
\end{figure}

We assess the impact of changepoints on seasonal trend estimates by comparing results from a GEV model that incorporates changepoints with those from a GEV model that does not. Figure~\ref{f:Scatter_Strends} summarizes this comparison for 423 stations where at least one changepoint was detected. Across all seasons, the two approaches show weak agreement, with only a very weak relationship between trend estimates. Incorporating changepoints alters not only the magnitude of estimated trends but, in some cases, also reverses their direction from positive to negative or vice versa. Notably, between 90 and 97 of the 423 stations (21.3--22.9\%, depending on season) exhibit seasonal trend differences exceeding 10 mm day$^{-1}$ century$^{-1}$. Furthermore, including changepoints increases the variability of seasonal trends, indicating that local changes in extreme precipitation may be more pronounced than previously found. Overall, these findings reinforce the importance of accounting for changepoints to obtain accurate trend estimates.

\begin{figure}[!ht]
\vspace{-5mm}
\centerline{\includegraphics[width=0.54\linewidth]{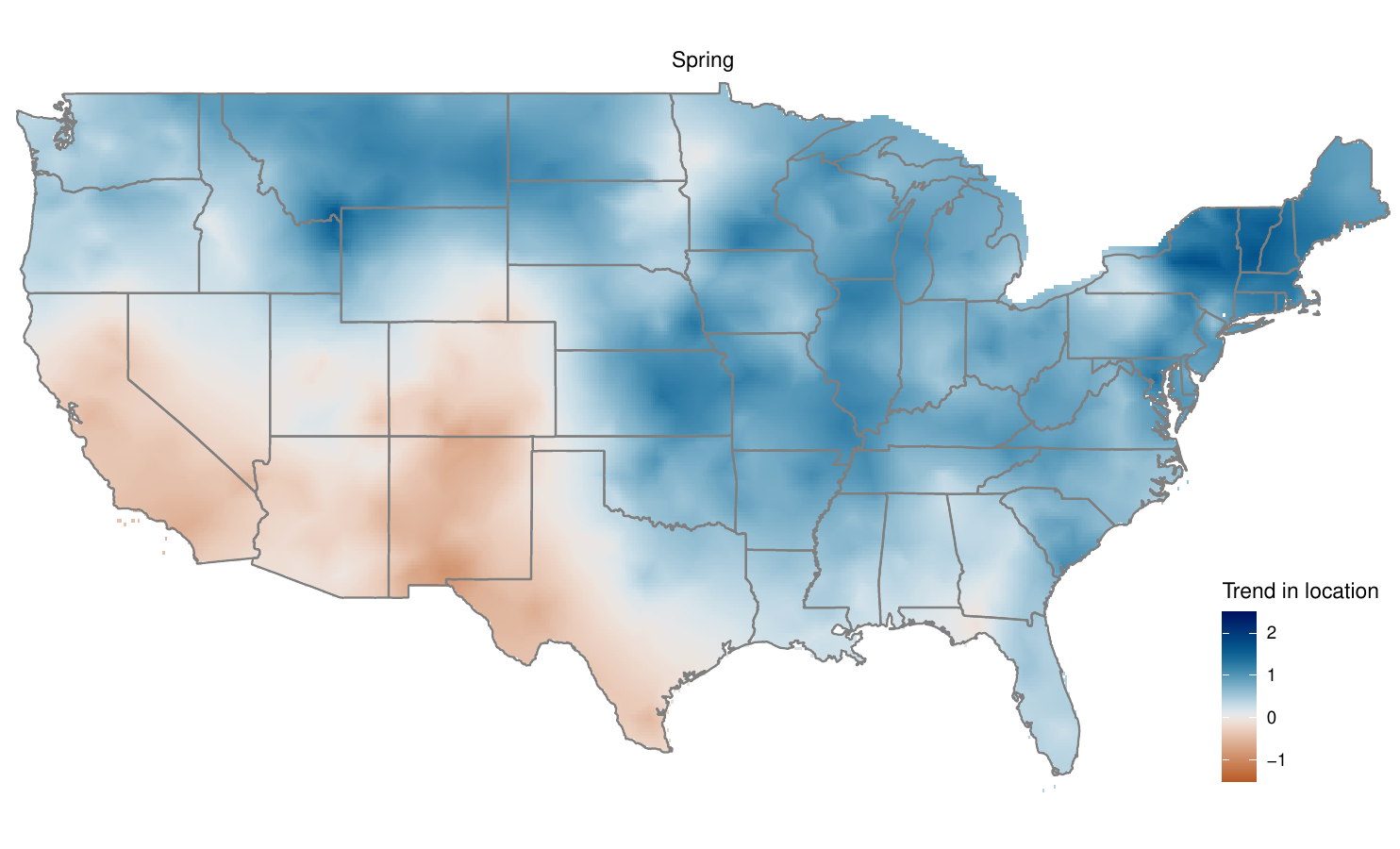}\hspace{-1mm}
            \includegraphics[width=0.54\linewidth]{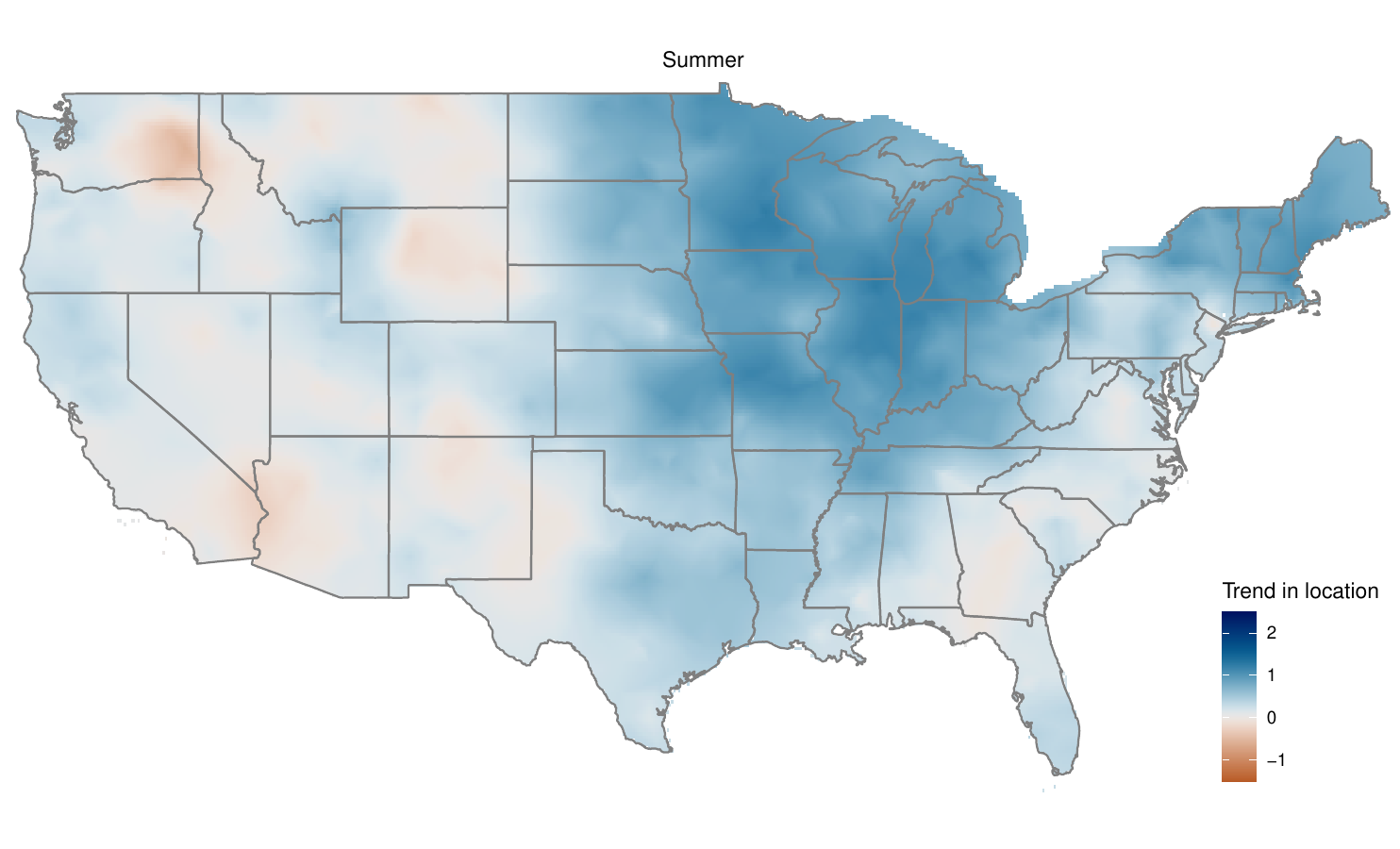}}\vspace{-3mm}
\centerline{\includegraphics[width=0.54\linewidth]{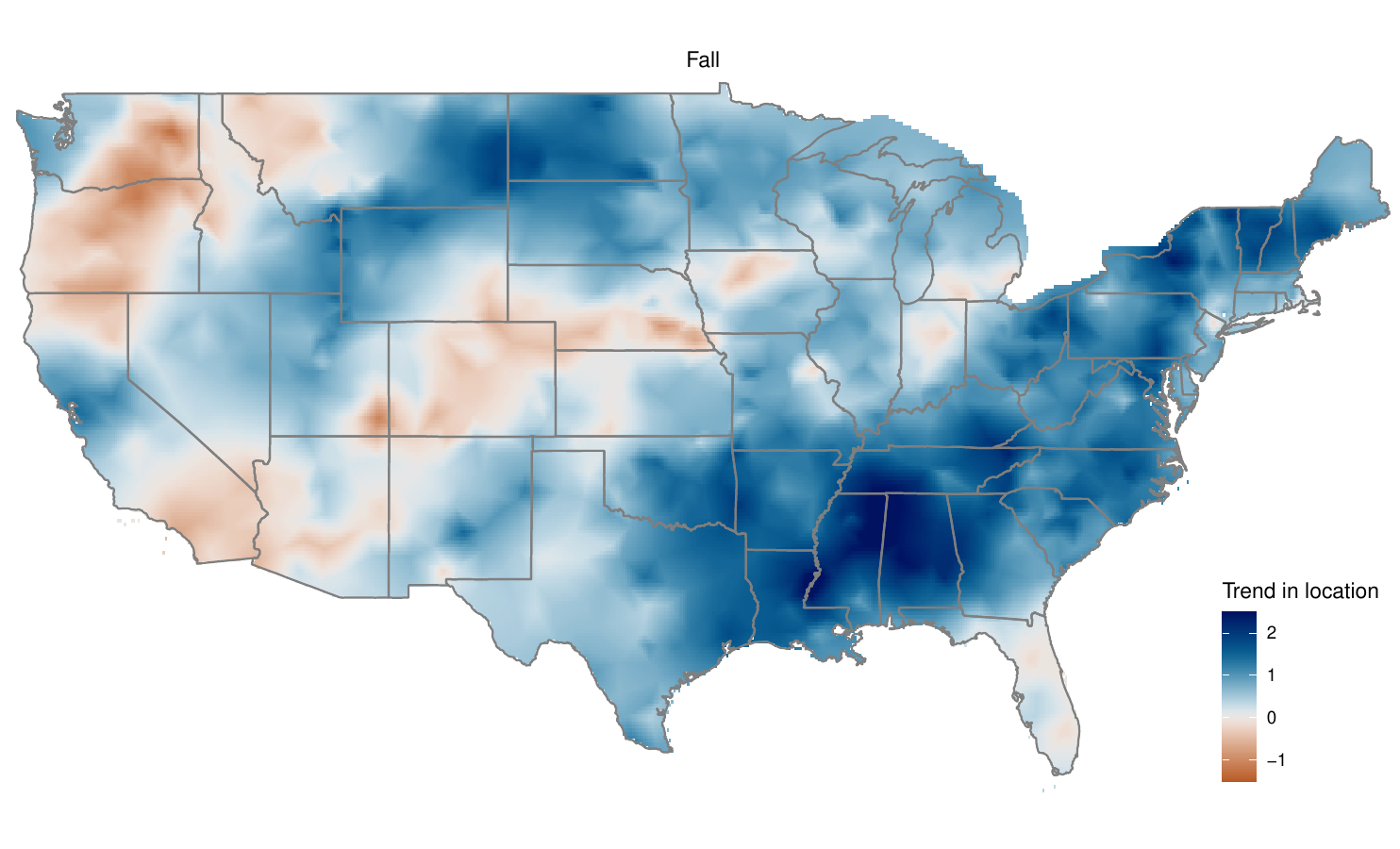}\hspace{-1mm}
            \includegraphics[width=0.54\linewidth]{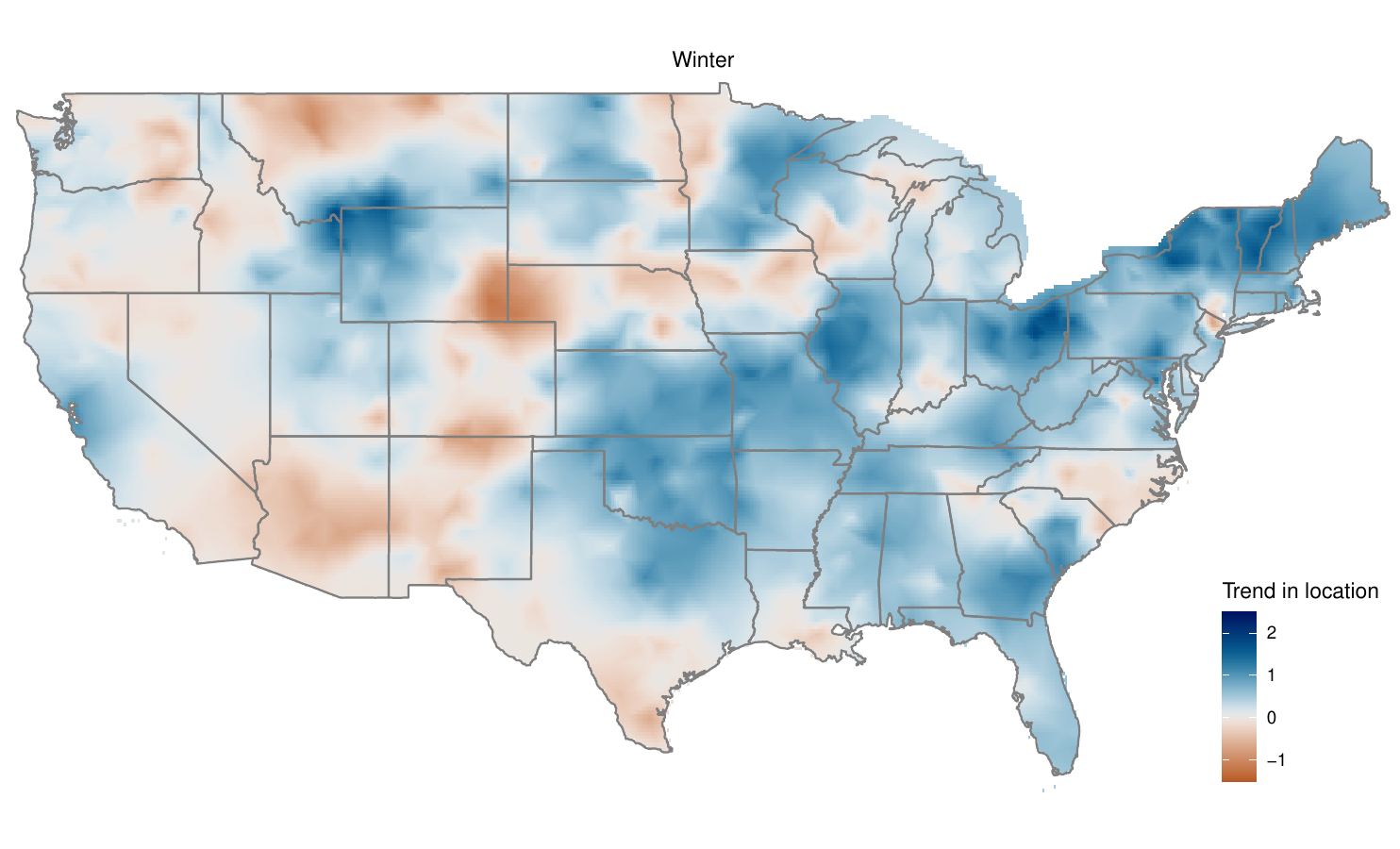}}
\vspace{-3mm}
\caption{Smoothed $Z$-statistics of US seasonal maximum precipitation trends: Spring (top left), Summer (top right), Fall (bottom left), and Winter (bottom right). Darker blue regions indicate more significantly increasing trends, while darker red regions signify more significantly decreasing trends.}
\label{f:STrendsZscore_US}
\end{figure}

To account for uncertainty in our seasonal trend estimates, we compute the season $s$ $Z$-statistic: $Z_{s} = \hat{\beta}_{1,s} / SE( \hat{\beta}_{1,s} )$, where the uncertainty of estimated seasonal trends, $SE( \hat{\beta}_{1,s} ) = \widehat{Var}( \hat{\beta}_{1,s} )^{1/2}$, is approximated using the inverse of the Hessian matrix from the likelihood-based optimization process as illustrated in Section~\ref{s:GA}. The $Z_{s}$ values are calculated from 1,057 locations and then smoothed over the contiguous US for each season. A value of $Z_{s}$ exceeding 1.96 or falling below --1.96 indicates a statistically significant increase or decrease at 5\% level, respectively, in season $s$ maximum precipitation. Figure~\ref{f:STrendsZscore_US} highlights regional and seasonal variations in significance. In spring, significance is generally weak, with notable increases in southern Montana--northwestern Wyoming and New England, while mild decreases appear in the Southwest. Summer trends are largely insignificant, though mild positive trends appear in the Midwest and New England, while mild negative trends occur in eastern Washington. In fall, significant increases emerge from the Southeast through the Northeast, as well as from the San Francisco Bay Area to eastern Idaho and North Dakota, whereas precipitation nearly significantly decreases in southern Washington, Oregon, and northern California. Winter trends show significant increases in southern Montana--northwestern Wyoming and New England, while decreases are observed in southwestern Wyoming, western Nebraska, and northern Montana. Central Arizona and eastern Washington also exhibit decreasing trends, though not significant at the 5\% level.

\begin{figure}[!ht]
\centerline{\includegraphics[width=0.68\linewidth]{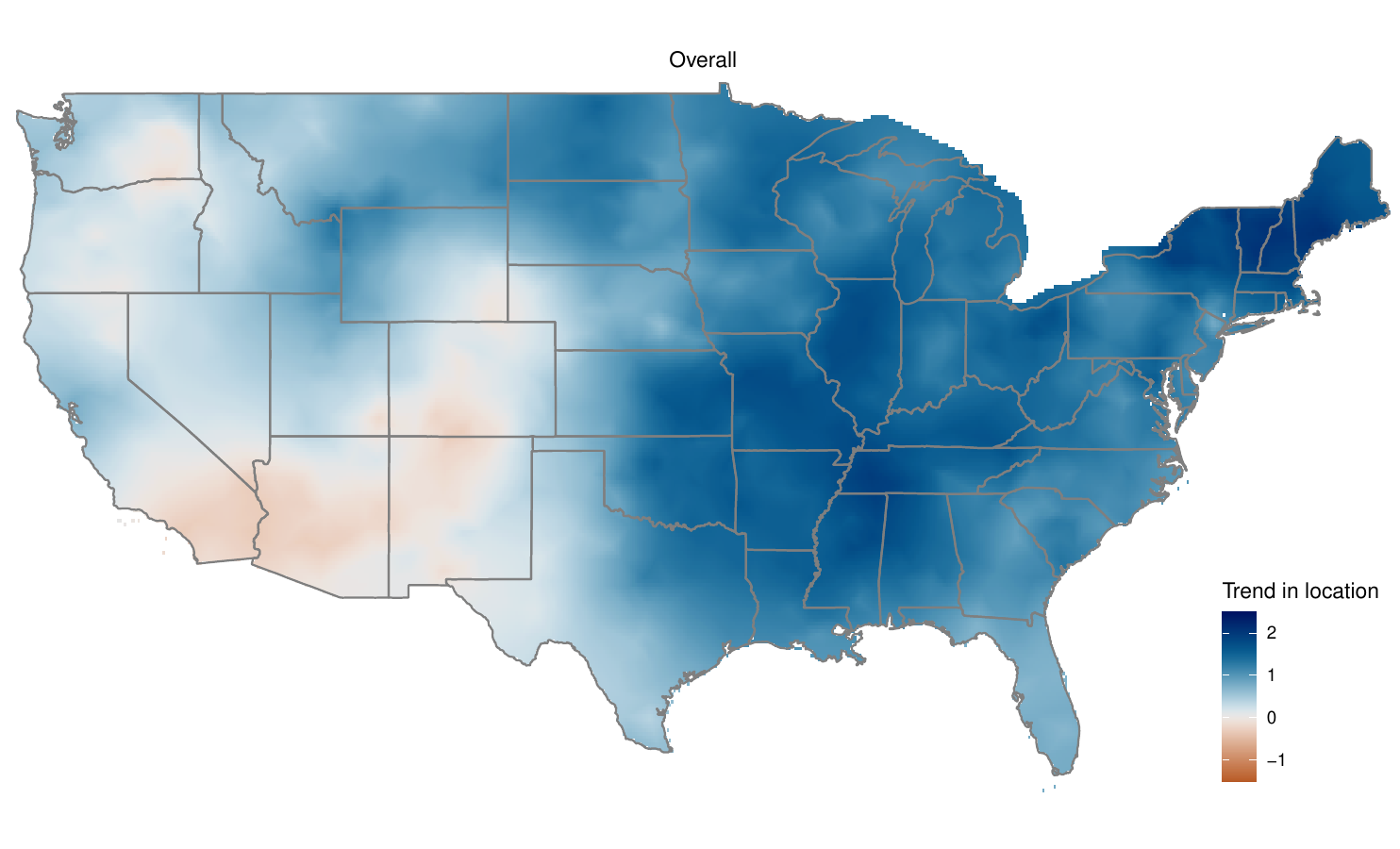}}
\vspace{-4mm}
\caption{Smoothed $Z$-statistics of US annual seasonal maximum precipitation average trends. Darker blue regions indicate more significantly increasing trends, while darker red regions signify more significantly decreasing trends.}
\label{f:All_TrendsZscore_US}
\end{figure}

Next, we assess the significance of the long-term trend in seasonal maximum precipitation averages by computing the $Z$-statistic: $Z_{\scriptscriptstyle\text{LTA}} = \hat{\beta}_{\scriptscriptstyle\text{LTA}} / SE( \hat{\beta}_{\scriptscriptstyle\text{LTA}} )$, as described in Section~\ref{s:LT_rl_intrp}, using data from 1,057 stations. Figure~\ref{f:All_TrendsZscore_US} illustrates the spatial distribution of these smoothed $Z_{\scriptscriptstyle\text{LTA}}$ values. A significantly increasing trend is evident in New England, with nearly significant positive trends in the Central Plains, while mild decreasing trends appear in the Southwest, southern Colorado, and eastern Washington.

The seasonal trend maps from this study reveal a greater prevalence of increasing maximum precipitation trends in the East, with the highest occurrence observed in the fall season. These upward trends in fall are most prominent in the Southern and Eastern regions. In contrast, the Southwest exhibits more decreasing trends across all seasons, with the most pronounced declines occurring in winter. In addition, greater variabilities have occurred in the South and East during the fall months and in the Central US throughout the winter. Based on these findings, further analysis will need to be conducted on those regions with significant changes in extreme precipitation.

The patterns of increasing trend in fall over the Gulf Coast states and along the East Coast appear to be related with hurricanes and tropical cyclones, due to the combination of warm ocean temperatures, atmospheric instability, and favorable wind patterns, as previously noted in literature \citep[cf.][]{Ashouri:others:2016}. Hurricanes contribute to extreme precipitation, flooding, and coastal erosion, which can alter river discharge, groundwater recharge, and soil moisture dynamics. The influx of stormwater often overwhelms drainage systems, leading to flash floods and prolonged inundation in urban and rural areas. Additionally, excessive rainfall from hurricanes can impact water quality, increasing sediment loads and pollutants in rivers and reservoirs. As hurricane activity intensifies due to climate change, understanding these hydrological changes is essential for improving flood forecasting, water resource management, and infrastructure resilience.

Accurate quantification of trends is essential for understanding past changes in extreme precipitation and predicting future changes. To achieve this, a rigorous and efficient method for detecting changepoints must be applied. Our analysis carefully accounts for changepoints and the diverse seasonal dynamics of extreme precipitation, including trends, seasonalities, variabilities, and return levels, to provide a comprehensive understanding of how extreme precipitation patterns evolve over time in the US. Recognizing that the trend and variability of extreme precipitation are not uniform across seasons or regions, we identify distinct spatial and temporal trends and variations. These seasonal and regional differences highlight the complex nature of extreme precipitation, emphasizing the need for a nuanced approach that captures both long-term trends and abrupt shifts in climate behavior. By incorporating changepoints and seasonal trend and variability into our analysis, we offer a more detailed and accurate assessment of extreme precipitation dynamics. Accounting for these changes could be valuable in preparing for weather events that may have serious physical and financial impacts on communities in these regions, and is also critical for improving risk assessment and adaptation strategies for our society. Moreover, a better understanding of heavy precipitation seasonality could enhance future water management and guide the allocation of water resources for agriculture, urban needs, and ecosystems \citep{Pal:others:2013}.

\section*{Acknowledgments}

The authors acknowledge that the computational analyses were run on Northern Arizona University's Monsoon computing cluster, funded by Arizona's Technology \& Research Initiative Fund (TRIF). Jaechoul Lee’s research was also supported by Northern Arizona University's TRIF. The authors would like to thank Meghan Edgerton for providing feedback on a preliminary version of the manuscript.

\spacingset{1.24}

\bibliographystyle{apalike}
\bibliography{precip_references}

\end{document}